\documentclass[aps,prl,10pt,letterpaper,twocolumn]{revtex4-1}

\usepackage{soul}
\usepackage{newtxtext, newtxmath}
\usepackage{xcolor, graphicx, hyperref}
\usepackage[T1]{fontenc}
\hypersetup{
    colorlinks,
    linkcolor={blue},
    citecolor={blue},
    urlcolor={blue},
    pdfauthor={David K. Tuckett, Stephen D. Bartlett, Steven T. Flammia, Benjamin J. Brown},
    pdftitle={Fault-tolerant thresholds for the surface code in excess of 5\% under biased noise}
}

\usepackage{amsmath,amsfonts,amssymb}

\usepackage[caption=false]{subfig}
\graphicspath{ {./fig/} }

\usepackage{algorithm2e}
\SetAlFnt{\normalsize}

\SetAlgoCaptionSeparator{.}
\SetAlCapNameFnt{\small}
\SetAlCapFnt{\small}

\begin{document}

\title{Fault-tolerant thresholds for the surface code in excess of 5\% under biased noise}

\author{David K. Tuckett}
\author{Stephen D. Bartlett}
\author{Steven T. Flammia}
\author{Benjamin J. Brown}
\affiliation{Centre for Engineered Quantum Systems, School of Physics, University of Sydney, Sydney, New South Wales 2006, Australia}

\date{\today}
\begin{abstract}
Noise in quantum computing is countered with quantum error correction. 
Achieving optimal performance will require tailoring codes and decoding algorithms to account for features of realistic noise, such as the common situation where the noise is biased towards dephasing. 
Here we introduce an efficient high-threshold decoder for a noise-tailored surface code based on minimum-weight perfect matching. 
The decoder exploits the symmetries of its syndrome under the action of biased noise and generalizes to the fault-tolerant regime where measurements are unreliable. 
Using this decoder, we obtain fault-tolerant thresholds in excess of $6\%$ for a phenomenological noise model in the limit where dephasing dominates. 
These gains persist even for modest noise biases: we find a threshold of $\sim 5\%$ in an experimentally relevant regime where dephasing errors occur at a rate 100 times greater than bit-flip errors. 
\end{abstract}

\maketitle

The surface code~\cite{Kitaev03, Dennis02} is among the most promising quantum error-correcting codes to realize the first generation of scalable quantum computers~\cite{Terhal15, Brown16, Campbell17}. 
This is due to its two-dimensional layout and low-weight stabilizers that help give it its high threshold~\cite{Dennis02, Wang03, Raussendorf07}, and its universal set of fault-tolerant logical gates~\cite{Dennis02, Bravyi05, Horsman12, Brown17, Brown19a}. 
Ongoing experimental work~\cite{Barends14, Corcoles15, Kelly15, Takita16} is steadily improving the surface code error rates. 
Concurrent work on improved decoding algorithms~\cite{Wang03, Raussendorf07, DuclosCianci10, Fowler12a, Bravyi14, Darmawan18} is leading to higher thresholds and lower logical failure rates, reducing the exquisite control demanded of experimentalists to realize such a system.

Identifying the best decoder for the surface code depends critically on the noise model. 
Minimum-weight perfect matching (MWPM)~\cite{Edmonds65, Kolmogorov09} is near optimal in the case of the standard surface code with a bit-flip error model~\cite{Dennis02} and for a phenomenological error model with unreliable measurements~\cite{Wang03}; see~\cite{Ohno04, Kubica17}.  
More recently, attention has turned to tailoring the decoder to perform under more realistic types of noise, such as depolarizing noise~\cite{DuclosCianci10, Wootton12, Fowler13, Bravyi14, Darmawan18} and correlated errors~\cite{Hutter14a, Nickerson17, Fowler12b}. 
Of particular note is noise that is biased towards dephasing: a common feature of many architectures. 
With biased noise and reliable measurements, it is known that the surface code can be tailored to accentuate commonly occurring errors such that an appropriate decoder will give substantially increased thresholds~\cite{Tuckett18, Tuckett18a}. 
However, these high thresholds were obtained using decoders with no known efficient implementation in the realistic setting where measurements are unreliable and the noise bias is finite.

In this Letter we propose a practical and efficient decoder that performs well for both finite bias and noisy measurements, demonstrating that the exceptional gains of the tailored surface code under biased noise extend to the fault-tolerant regime.
We use the MWPM algorithm together with a recent technique to exploit symmetries of a given quantum error-correcting code~\cite{Brown19}. 
Rather than using the symmetries of the code, we generalize this idea and use the symmetries of the entire system. 
Specifically, we exploit the symmetries of the syndrome \textit{with respect to its incident error model}.
Applied to pure dephasing noise, our decoder exploits the one-dimensional symmetries of the system by pairing the defects of each symmetry separately.
Crucially, our approach readily extends to the situation where measurements are unreliable, as well as the finite-bias regime where some low-rate errors violate the symmetries we rely on. 
We demonstrate that our approach leads to fault-tolerant thresholds exceeding $6\%$ for infinite bias, with these substantial gains persisting to modest biases. 
Comparing with the \emph{optimal} threshold of $3.3\%$~\cite{Ohno04, Kubica17} for conventional decoders that correct the bit-flip and dephasing errors of the same noise model separately, our results represent a very significant improvement in the level of noise that can be tolerated in practical quantum technologies.

\begin{figure}
\includegraphics{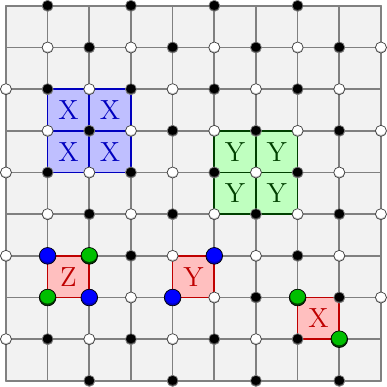}  \quad  \includegraphics{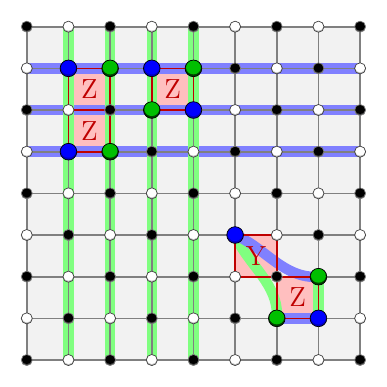}
\caption{\label{Fig:Lattice} (Left) The surface code with qubits on the faces of a square $d \times d$ lattice. The vertices $v$ are bicolored such that stabilizer generators $S_v = \prod_{ \partial f \ni v} X_f$ ($S_v = \prod_{\partial f \ni v }Y_f$) lie on black (white) vertices, and $\partial f \ni v$ denotes the faces $f$ touching $v$. Examples are shown at the top of the figure. 
The syndrome patterns for Pauli $X$, $Y$ and $Z$ errors are shown at the bottom of the figure. 
(Right) The surface code with periodic boundary conditions.  Our noise model is such that $Z$ errors occur at a higher rate than Pauli $X$ or $Y$ errors. 
The syndromes of $Z$ errors, shown at the top left of the figure, respect one-dimensional symmetries, shown as blue and green lines. 
We can therefore consistently match vertices along the rows and columns of the lattice. 
The edges returned from each MWPM subroutine reproduce the boundary of the faces that support the error. 
Lower-rate nondephasing errors may violate the symmetries of the system (bottom, right).
}
\end{figure}

\paragraph{Surface code tailored for dephasing.} We define the surface code in a rotated basis with $X$- and $Y$-type stabilizers, $S_v \in \mathcal{S}$, to provide additional syndrome information about $Z$ errors; see Fig.~\ref{Fig:Lattice} and its corresponding caption.
We consider errors $E \in \mathcal{E}$ drawn from a subgroup of the Pauli group $ \mathcal{E} \subseteq \mathcal{P} $. 
We define the syndrome as a list of the locations of defects. 
For a given error, defects lie on vertices $v$ such that $S_v E | \psi \rangle =  (-1) E |\psi \rangle$ for code states $|\psi \rangle$ satisfying $S_v |\psi \rangle = |\psi \rangle $ for all $v$.

\paragraph{Decoding with symmetry.} We first consider the infinite bias (pure-dephasing) error model generated by only $Z$ errors, $ \mathcal{E}^Z = \left\langle Z_f \right\rangle$. 
Errors drawn from this model respect one-dimensional symmetries of the lattice, as in Fig.~\ref{Fig:Lattice}. 
A single $Z$ error generates two defects on each of its adjacent rows and columns. 
Up to boundary conditions, any error drawn from $\mathcal{E}^Z$ will respect a defect parity conservation symmetry on each of the rows and columns of the lattice.

Let us make this notion of a symmetry rigorous. 
A symmetry is specified by a subgroup of the stabilizer group $\mathcal{S}_{\text{sym}} \subseteq \mathcal{S}$.  Elements $S \in \mathcal{S}_{\text{sym}}$ are defined with respect to an error model $\mathcal{E}$ such that they satisfy $S E |\psi \rangle = (+1) E |\psi \rangle $ for all $E \in \mathcal{E}$ and code states $|\psi \rangle$.
This generalizes Ref.~\cite{Brown19} where symmetries are defined for the special case where $\mathcal{E} = \mathcal{P}$; the symmetry is now a function of the combined system of both the code and the error model.

For general Pauli error models, the surface code has global symmetries~\cite{Kitaev03}; $\prod_{v\in\mathcal{G}}S_v = 1$ with $\mathcal{G}$ the set of either black or white vertices where we briefly assume periodic boundary conditions to illustrate this point. 
Under pure dephasing noise, the same model has a much richer set of one-dimensional symmetries. 
Observe that $S_{\mathcal{L}} = \prod_{v \in \mathcal{L}} S_v$, with $\mathcal{L}$ the set of vertices on a row or column, is a product of Pauli $Z$ matrices. 
As such, the one-dimensional stabilizers $S_\mathcal{L}$ commute with errors drawn from $\mathcal{E}^Z$ and are therefore symmetries. 
The set of all such $S_{\mathcal{L}}$ generate $\mathcal{S}_{\text{sym}}$ with respect to $\mathcal{E}^Z$.

Now consider what this symmetry implies for an arbitrary syndrome in our error model. 
A direct consequence of the definition of a symmetry $\mathcal{S}_{\text{sym}}$ is that, for pure dephasing noise, there will always be an even number of defects measured by the subsets of stabilizers whose product gives elements of $\mathcal{S}_{\text{sym}}$. 
We can design a decoder that exploits this property of these subsets. 
Specifically, we can consistently pair the defects detected by the stabilizers of these subsets using, say, MWPM, or another suitable pairing algorithm such as that of Ref.~\cite{Delfosse17}.  
Collections of defects that are combined with pairing operations on sufficiently many symmetries can be neutralized with a low-weight Pauli operator. We say that such a collection is locally correctable~\cite{Brown19}. 
For the surface code under pure dephasing noise, by performing pairing over the one-dimensional lattice symmetries, the edges returned from MWPM form the boundary of the error; see Fig.~\ref{Fig:Lattice}(right). 
The interior of the boundary determines the correction.

Such a decoder is readily extended to the fault-tolerant setting where measurements are unreliable and may give incorrect outcomes. 
A single measurement error will violate the defect symmetries of the two-dimensional system. 
Following the approach of Ref.~\cite{Dennis02}, we can recover a new symmetry in the fault-tolerant setting in $(2+1)$-dimensional spacetime by repeating stabilizer measurements over time, see also Ref.~\cite{Brown19}. 
A symmetry is recovered by taking the parity of pairs of sequential measurement outcomes, with odd parity heralding a defect. 
This spacetime symmetry is generic to our proposal here. 
In this situation, up to the lattice boundaries, the symmetries represent constraints among collections of defects lying on $(1+1)$-dimensional planes. 
Curiously, unlike the phenomenological bit-flip noise model for the surface code~\cite{Wang03, Raussendorf06}, the biased phenomenological error model considered here is anisotropic in spacetime.
We emphasize the importance of checking for temporal logical errors, consisting of strings of sequential measurement errors, as they may introduce logical failures while performing code deformations~\cite{Vuillot18}.

The symmetries of the system are altered at lattice boundaries.  
We can adapt the decoder to account for this, by adding a pair of defects at each time step to all vertices where a stabilizer is not imposed at the boundary; see Fig.~\ref{Fig:Lattice}(left). 
These defects can be paired to other defects within their respective $(1+1)$-dimensional planes of symmetry.
Otherwise, they can be matched together freely in the case that they do not need to be paired.

\paragraph{Decoding with finite bias.} We next adapt our decoder to deal with low-rate $X$ and $Y$ errors in addition to high-rate $Z$ errors. 
For simplicity we will describe this modification for the case of periodic boundary conditions and where measurements are reliable. We give a technical description of all the decoders we present in the Supplemental Material.

The decoder for infinite bias noise will pair each defect of the system twice: once to a horizontally separated defect and once to a vertically separated defect. 
Low rate $X$ and $Y$ errors violate the one-dimensional symmetries that enable us to use the  strategy described above, but we can weakly adhere to the strategy as follows. 
In our modified decoder we pair all defects twice: once where we strongly bias the decoder to pair defects horizontally, and a second time where we strongly bias each defect to pair vertically. 
Unlike in the infinite-bias case, we permit our decoder to pair defects that are not within their same row or column. 
We penalize such pairings according to the amount of noise bias.
This can be achieved in our input into the MWPM algorithm by assigning high weights to edges for pairs of defects that are not aligned on the same row or column, depending on the respective matching.

In the case of finite bias the collections of defects that are connected through the edges returned by pairing may not be locally correctable. 
We deal with this issue with additional use of MWPM to complete the decoding procedure. 
One can show that there will be an even number of defects in each collection of defects connected by edges. 
Therefore, the parity of defects on black and white vertices are equal. 
We call collections of defects with an even (odd) parity of defects on black and white vertices `neutral' (`charged'). 
Neutral clusters can be locally corrected. 
Remaining charged collections of defects can be made neutral by pairing them with other nearby charged collections of defects. 
This final pairing ensures the collections of connected defects are locally correctable.

\paragraph{Biased noise models.} We will test our decoder under two scenarios: with a biased noise model and ideal measurements, and with a phenomenological biased noise model with unreliable measurements.  
At each time step, qubits are subjected to an error with probability $p$.  
Pauli $Z$ errors occur at high rate $p_{\text{h.r.}} = p \eta / (\eta +1)$, while $X$ and $Y$ errors occur at a lower rate, $ p_{\text{l.r.}} = p / 2(\eta + 1)$. 
The phenomenological (ideal-measurement) biased noise model gives an incorrect measurement outcome with probability $q=p$ ($q=0$).

\begin{figure}
\includegraphics{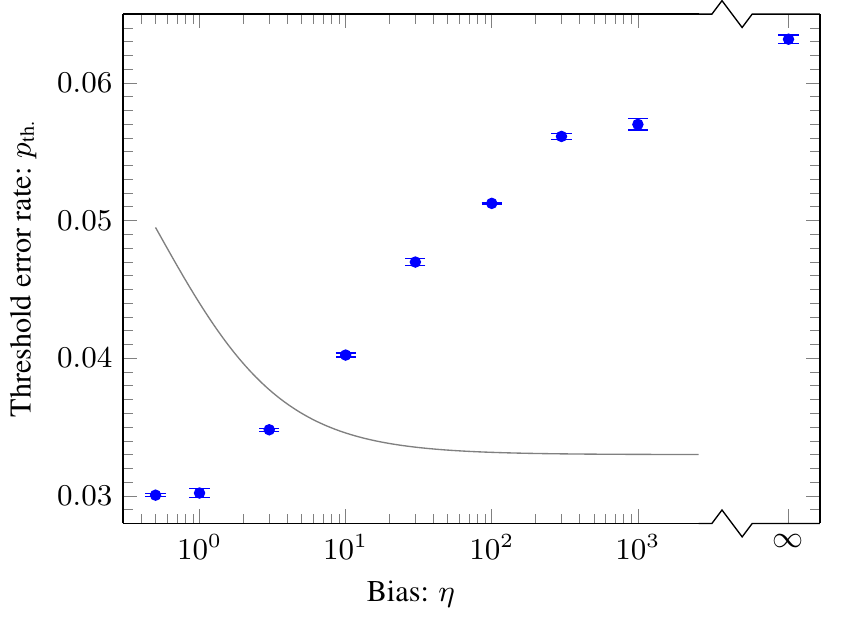}
\caption{\label{Fig:BiasNoiseFaultTolerant}
  Threshold error rates $p_{\text{th.}}$ as a function of noise bias $\eta$ for both spatial and temporal logical errors for the surface code with periodic boundary conditions.
  The points show threshold estimates together with 1 standard deviation error bars.
  The points at smallest and largest bias values correspond to $\eta=0.5$ (depolarizing noise), and $\eta=\infty$ (pure dephasing), respectively.
  The solid line represents the optimal performance for the standard surface code with phenomenological noise of a decoder that deals with bit-flip errors and dephasing noise separately.
  Codes with distance $d=12, 14, 16, 18, 20$ and $d=24, 28, 32, 36, 40$ were used for finite and infinite bias threshold estimates, respectively. 
}
%
\end{figure}

It is important to consider whether the phenomenological noise model we introduced is compatible with a noise-bias setting~\cite{Aliferis08}.  
As we now demonstrate, it is possible to measure stabilizers and maintain the bias.  
Following the standard approach~\cite{Dennis02}, stabilizers are measured by preparing an ancilla, $a$, in an eigenstate of $X$, then applying entangling gates between the ancilla and the qubits that support the stabilizer, and finally measuring the ancilla qubit in the $X$ basis. 
To measure $S_v$ for black vertices $v$ we apply $\prod_{\partial f \ni v} CX_{a,f}$ where $CX_{a,f} = (1 + Z_a + X_f -Z_a X_f)/2$ is the controlled-not gate.
To measure white vertex stabilizers, we replace the $CX_{a,f}$ gates with $CY_{a,f} $ gates. 
These gates differ by an $\exp(\text{i} \pi Z_f / 2)$ rotation.

We can now justify that stabilizer measurements performed this way preserve the noise bias. 
Specifically, we demonstrate that no steps in the stabilizer circuit cause high-rate errors to introduce $X$ or $Y$ errors to the data qubits of the surface code. 
The $CX_{a,f}$ commutes with $Z$ errors that act on the ancilla. 
As such, it will not create high rate $X$ or $Y$ errors on the data qubits. 
Similarly, the single-qubit rotation that maps $CX_{a,f}$ onto $CY_{a,f}$ commutes with the high-rate errors, and will therefore only map low-rate errors onto other low-rate errors. 
Ancilla qubits are vulnerable to high-rate Pauli $Z$ errors. 
This is reflected by the error model that has a high measurement error rate, $q=p$.  
An additional concern is that the entangling gates such as $CX_{a,f}$ may increase the frequency that low-rate errors occur. 
This will depend on the physical implementation, and recent proposals have demonstrated that noise-bias-preserving $CX_{a,f}$ gates are indeed possible in some architectures~\cite{Puri19}.

\paragraph{Numerical simulations.} We simulate the performance of our decoder for the surface code with periodic boundary conditions against the phenomenological biased noise model, using 30\,000 trials per code distance and physical error probability.
We used the critical exponent method of Ref.~\cite{Wang03}, fitting to a quadratic model, to obtain threshold estimates with jackknife resampling over code distances to determine error bounds.  Because of the anisotropy in spacetime, we might expect the thresholds of logical errors in the spatial and temporal direction to differ. 
We report a failure if a logical error occurs in either the spatial or temporal direction.  
Our results are shown in Fig.~\ref{Fig:BiasNoiseFaultTolerant}.  
We identify a threshold of $6.32(3)\%$ for pure dephasing, and thresholds of $\sim 5\%$ for biases around $\eta = 100$. 
Our decoder begins to outperform the optimal values for standard methods, where bit-flip and dephasing errors are corrected separately, at $\eta \sim 5$. 
These results demonstrate the advantage of using our decoder in the fault-tolerant setting, even if the noise bias is modest.

\begin{figure}
\includegraphics{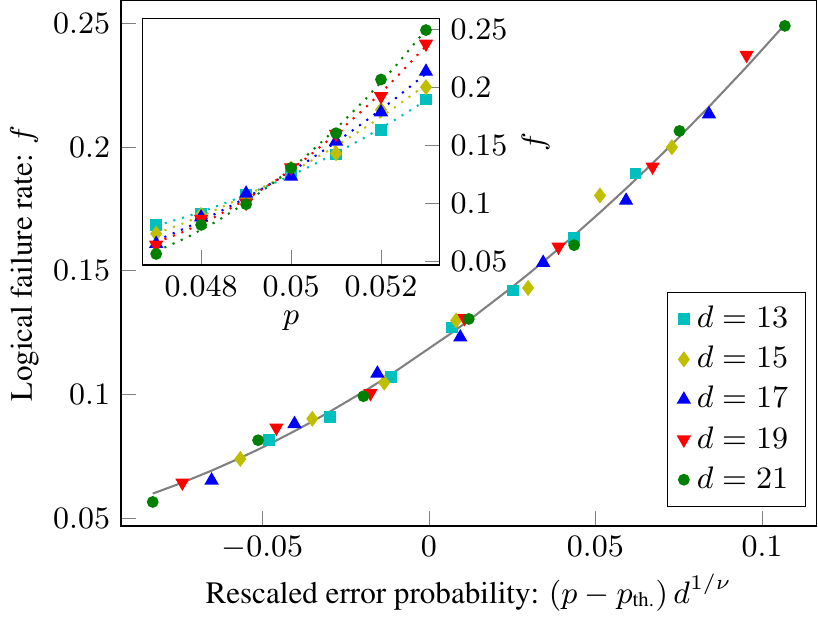}
\caption{\label{Fig:ThresholdPlot}
Numerical data demonstrating a finite threshold in the fault-tolerant setting. Logical (spatial) failure rate $f$ for the surface code with boundaries shown as a function of the rescaled error rate $x = (p-p_{\text{th.}}) d^{1/\nu}$ with bias $\eta = 100$ and $p_{\rm th.} = 4.96(1)\%$. 
  The solid line is the best fit to the model $f = A + Bx+Cx^2$. 
  The insets show the raw sample means over 30\,000 runs for various values of $p$.
}
\end{figure}

We have simulated the performance on the surface code with boundaries, yielding similar results. Figure~\ref{Fig:ThresholdPlot} demonstrates a threshold using the fault-tolerant decoder for the surface code with boundaries where $\eta=100$.
In this case we only measure spatial logical errors because there are no topologically nontrivial temporal errors. 
Remarkably, the threshold is very similar to the threshold obtained in the case with periodic boundary conditions where we also count logical failures along the temporal direction as well. 
This is surprising given the anisotropy of the decoding problem in the spatial and temporal directions.

\begin{figure}
\includegraphics{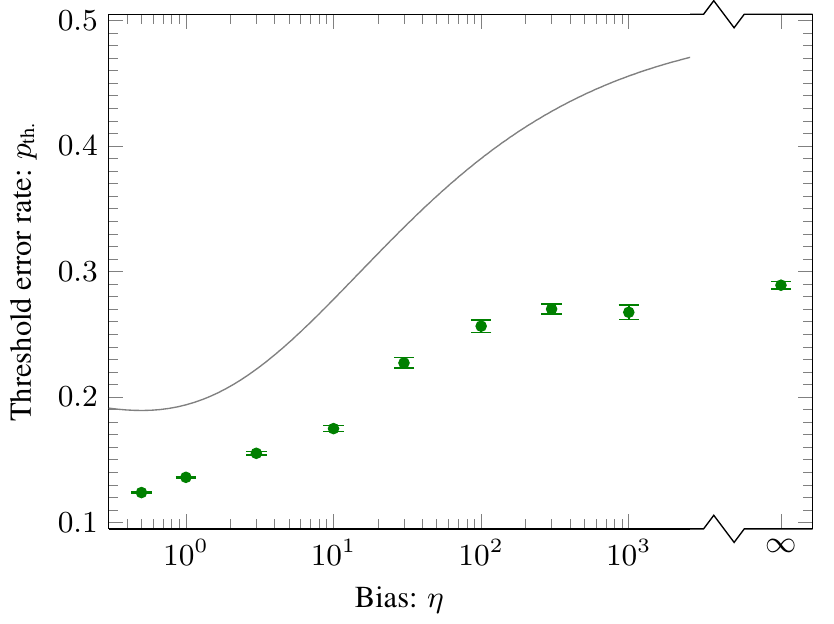}
\caption{\label{Fig:BiasNoiseIdeal}
  Threshold error rates $p_{\text{th.}}$ as a function of noise bias $\eta$ for the surface code with periodic boundary conditions and ideal measurements.
  The points show threshold estimates with 1 standard deviation error bars.
  The points at smallest and largest bias values correspond to $\eta=0.5$ (depolarizing noise), and $\eta=\infty$ (pure dephasing), respectively.
  The solid line, which is the zero-rate hashing bound for the associated Pauli error channel, represents threshold error rates that are achievable with ML decoding~\cite{Tuckett18a}.
  Codes with distance $d=24, 28, 32, 36, 40$ and $d=48, 56, 64, 72, 80$ were used for finite and infinite bias threshold estimates, respectively.
}
%
\end{figure}

We benchmark our decoder against the optimal performance of the surface code under the biased noise model. 
In the absence of optimal fault-tolerant thresholds (say, from statistical mechanical arguments~\cite{Chubb18}), we benchmark using the ideal measurement model.  In this case, optimal performance corresponds to the zero-rate hashing bound, which is achievable using a ML decoder~\cite{Bravyi14, Tuckett18a}.  We see in Fig.~\ref{Fig:BiasNoiseIdeal} that our decoder underperforms in comparison to the ML decoder, suggesting that there is considerable scope for further improvements.  A natural proposal would be to incorporate belief propagation into the MWPM algorithm.  Choices of boundary conditions also play a role.
We note that our decoder applied to the surface code with boundaries can achieve the optimal threshold of $p_{\text{th.}} \sim 1/2$ for pure dephasing noise.  However it underperforms similarly to that shown in Fig.~\ref{Fig:BiasNoiseIdeal} at finite biases.

\paragraph{Low error rates.} The performance of the decoder below threshold will determine the resources required to perform quantum computation. 
We now speculate on the logical failure rates where the physical error rate is low, specifically $p \ll 1 / d$. 
Using conventional decoding methods the logical failure rate decays as $\mathcal{O}( p^{ \delta \sqrt{n}} )$~\cite{Dennis02, Beverland18} with $n = d\times d$ the code length and $\delta$ a constant. 
The high-threshold at infinite bias is indicative that the decoder can tolerate up to $\sim n / 2$ dephasing errors ~\cite{Tuckett18,Tuckett18a}. 
We may therefore expect that the logical failure rate will decay with improved scaling, $\mathcal{O}( {p_{\text{h.r.}}} ^{\alpha n} )$, for some constant $\alpha$.

At finite noise bias, the improved scaling in logical failure rate with $n$ can only persist up to some critical system size.
Above some system size that depends on $\eta$, we expect that the most likely error that will cause a logical failure will be due a string consisting of $\sim \sqrt{n}$ low-rate errors. 
Up to constant factors, this will occur for some $n$ where ${p_{\text{h.r.}}}^{\alpha n}  \ll  {p_{\text{l.r.}}}^{\delta\sqrt{n}}$. 
Nevertheless, given high bias, the decoder will vastly improve logical error rates in the regime where the most likely failure mechanisms are due to long strings of low-rate error events.

We contrast this with bias nullification schemes by concatenation~\cite{Stephens13,Xu18}. These approaches increase the effective rate that uncommon errors act on the surface code by a factor equal to the number of qubits of each repetition code, leading to worse performance at low error rates. 
Moreover, they can only tolerate at most $\propto \sqrt{n}$ high-rate errors.

Extending again to the fault-tolerant case, temporal-logical errors are caused by strings of measurement errors that occur at a high rate. 
We should consider increasing the number of repetitions $T$ of the error-correction cycle between code deformations to reduce the likelihood of temporal logical failures. 
Choosing $T \sim 2 \delta \sqrt{n} \log p_{\text{l.r.}} / \log p$ will ensure temporal errors will occur at a rate similar to spatial logical errors, $\sim {p_{\text{l.r.}}}^{\delta\sqrt{n}}$, where we have assumed a temporal logical error occurs with likelihood $\sim p^{T/2}$. 
To achieve the target logical failure rate of the system, although the qubits will be occupied for a longer time to decrease the failure rate of temporal logical errors, the associated decrease in the two spatial dimensions will result in a net improvement on resource scaling using our system.

\paragraph{Discussion.} 
Minimum-weight perfect matching has formed the backbone of topological quantum error correction~\cite{Dennis02, Wang03, Raussendorf06, Brown19, Fowler12a, Delfosse14, Kubica19}. 
The realization that we can design MWPM decoders with knowledge of the symmetries of the code or system opens up a number of new avenues for decoding algorithm design. 
A multitude of codes have yet to be explored, as well as their interaction with specialized noise models that reflect the errors that occur in the laboratory.  Significant improvements in fault-tolerant thresholds obtained though tailored codes and realistic noise models, such as those we have demonstrated here, offer great promise for the realization of practical quantum technologies.

\begin{acknowledgments}
We are grateful for helpful and supportive conversations with A. Grimsmo, N. Nickerson and D. Williamson. 
This work is supported by the Australian Research Council via the Centre of Excellence in Engineered Quantum Systems (EQUS) project number CE170100009. 
BJB is also supported by the University of Sydney Fellowship Programme.
Access to high-performance computing resources was provided by the National Computational Infrastructure (NCI), which is supported by the Australian Government, and by the Sydney Informatics Hub, which is funded by the University of Sydney.
\end{acknowledgments}

\bibstyle{plain}


\begin{thebibliography}{42}%
\makeatletter
\providecommand \@ifxundefined [1]{%
 \@ifx{#1\undefined}
}%
\providecommand \@ifnum [1]{%
 \ifnum #1\expandafter \@firstoftwo
 \else \expandafter \@secondoftwo
 \fi
}%
\providecommand \@ifx [1]{%
 \ifx #1\expandafter \@firstoftwo
 \else \expandafter \@secondoftwo
 \fi
}%
\providecommand \natexlab [1]{#1}%
\providecommand \enquote  [1]{``#1''}%
\providecommand \bibnamefont  [1]{#1}%
\providecommand \bibfnamefont [1]{#1}%
\providecommand \citenamefont [1]{#1}%
\providecommand \href@noop [0]{\@secondoftwo}%
\providecommand \href [0]{\begingroup \@sanitize@url \@href}%
\providecommand \@href[1]{\@@startlink{#1}\@@href}%
\providecommand \@@href[1]{\endgroup#1\@@endlink}%
\providecommand \@sanitize@url [0]{\catcode `\\12\catcode `\$12\catcode
  `\&12\catcode `\#12\catcode `\^12\catcode `\_12\catcode `\%12\relax}%
\providecommand \@@startlink[1]{}%
\providecommand \@@endlink[0]{}%
\providecommand \url  [0]{\begingroup\@sanitize@url \@url }%
\providecommand \@url [1]{\endgroup\@href {#1}{\urlprefix }}%
\providecommand \urlprefix  [0]{URL }%
\providecommand \Eprint [0]{\href }%
\providecommand \doibase [0]{http://dx.doi.org/}%
\providecommand \selectlanguage [0]{\@gobble}%
\providecommand \bibinfo  [0]{\@secondoftwo}%
\providecommand \bibfield  [0]{\@secondoftwo}%
\providecommand \translation [1]{[#1]}%
\providecommand \BibitemOpen [0]{}%
\providecommand \bibitemStop [0]{}%
\providecommand \bibitemNoStop [0]{.\EOS\space}%
\providecommand \EOS [0]{\spacefactor3000\relax}%
\providecommand \BibitemShut  [1]{\csname bibitem#1\endcsname}%
\let\auto@bib@innerbib\@empty
\bibitem [{\citenamefont {Kitaev}(2003)}]{Kitaev03}%
  \BibitemOpen
  \bibfield  {author} {\bibinfo {author} {\bibfnamefont {A.~Y.}\ \bibnamefont
  {Kitaev}},\ }\href@noop {} {\bibfield  {journal} {\bibinfo  {journal} {Ann.
  Phys.}\ }\textbf {\bibinfo {volume} {303}},\ \bibinfo {pages} {2} (\bibinfo
  {year} {2003})}\BibitemShut {NoStop}%
\bibitem [{\citenamefont {Dennis}\ \emph {et~al.}(2002)\citenamefont {Dennis},
  \citenamefont {Kitaev}, \citenamefont {Landahl},\ and\ \citenamefont
  {Preskill}}]{Dennis02}%
  \BibitemOpen
  \bibfield  {author} {\bibinfo {author} {\bibfnamefont {E.}~\bibnamefont
  {Dennis}}, \bibinfo {author} {\bibfnamefont {A.}~\bibnamefont {Kitaev}},
  \bibinfo {author} {\bibfnamefont {A.}~\bibnamefont {Landahl}}, \ and\
  \bibinfo {author} {\bibfnamefont {J.}~\bibnamefont {Preskill}},\ }\href@noop
  {} {\bibfield  {journal} {\bibinfo  {journal} {J. Math. Phys.}\ }\textbf
  {\bibinfo {volume} {43}},\ \bibinfo {pages} {4452} (\bibinfo {year}
  {2002})}\BibitemShut {NoStop}%
\bibitem [{\citenamefont {Terhal}(2015)}]{Terhal15}%
  \BibitemOpen
  \bibfield  {author} {\bibinfo {author} {\bibfnamefont {B.~M.}\ \bibnamefont
  {Terhal}},\ }\href@noop {} {\bibfield  {journal} {\bibinfo  {journal} {Rev.
  Mod. Phys.}\ }\textbf {\bibinfo {volume} {87}},\ \bibinfo {pages} {307}
  (\bibinfo {year} {2015})}\BibitemShut {NoStop}%
\bibitem [{\citenamefont {Brown}\ \emph {et~al.}(2016)\citenamefont {Brown},
  \citenamefont {Loss}, \citenamefont {Pachos}, \citenamefont {Self},\ and\
  \citenamefont {Wootton}}]{Brown16}%
  \BibitemOpen
  \bibfield  {author} {\bibinfo {author} {\bibfnamefont {B.~J.}\ \bibnamefont
  {Brown}}, \bibinfo {author} {\bibfnamefont {D.}~\bibnamefont {Loss}},
  \bibinfo {author} {\bibfnamefont {J.~K.}\ \bibnamefont {Pachos}}, \bibinfo
  {author} {\bibfnamefont {C.~N.}\ \bibnamefont {Self}}, \ and\ \bibinfo
  {author} {\bibfnamefont {J.~R.}\ \bibnamefont {Wootton}},\ }\href@noop {}
  {\bibfield  {journal} {\bibinfo  {journal} {Rev. Mod. Phys.}\ }\textbf
  {\bibinfo {volume} {88}},\ \bibinfo {pages} {045005} (\bibinfo {year}
  {2016})}\BibitemShut {NoStop}%
\bibitem [{\citenamefont {Campbell}\ \emph {et~al.}(2017)\citenamefont
  {Campbell}, \citenamefont {Terhal},\ and\ \citenamefont
  {Vuillot}}]{Campbell17}%
  \BibitemOpen
  \bibfield  {author} {\bibinfo {author} {\bibfnamefont {E.~T.}\ \bibnamefont
  {Campbell}}, \bibinfo {author} {\bibfnamefont {B.~M.}\ \bibnamefont
  {Terhal}}, \ and\ \bibinfo {author} {\bibfnamefont {C.}~\bibnamefont
  {Vuillot}},\ }\href@noop {} {\bibfield  {journal} {\bibinfo  {journal}
  {Nature}\ }\textbf {\bibinfo {volume} {549}},\ \bibinfo {pages} {172}
  (\bibinfo {year} {2017})}\BibitemShut {NoStop}%
\bibitem [{\citenamefont {Wang}\ \emph {et~al.}(2003)\citenamefont {Wang},
  \citenamefont {Harrington},\ and\ \citenamefont {Preskill}}]{Wang03}%
  \BibitemOpen
  \bibfield  {author} {\bibinfo {author} {\bibfnamefont {C.}~\bibnamefont
  {Wang}}, \bibinfo {author} {\bibfnamefont {J.}~\bibnamefont {Harrington}}, \
  and\ \bibinfo {author} {\bibfnamefont {J.}~\bibnamefont {Preskill}},\
  }\href@noop {} {\bibfield  {journal} {\bibinfo  {journal} {Ann. Phys.}\
  }\textbf {\bibinfo {volume} {303}},\ \bibinfo {pages} {31} (\bibinfo {year}
  {2003})}\BibitemShut {NoStop}%
\bibitem [{\citenamefont {Raussendorf}\ and\ \citenamefont
  {Harrington}(2007)}]{Raussendorf07}%
  \BibitemOpen
  \bibfield  {author} {\bibinfo {author} {\bibfnamefont {R.}~\bibnamefont
  {Raussendorf}}\ and\ \bibinfo {author} {\bibfnamefont {J.}~\bibnamefont
  {Harrington}},\ }\href@noop {} {\bibfield  {journal} {\bibinfo  {journal}
  {Phys. Rev. Lett.}\ }\textbf {\bibinfo {volume} {98}},\ \bibinfo {pages}
  {190504} (\bibinfo {year} {2007})}\BibitemShut {NoStop}%
\bibitem [{\citenamefont {Bravyi}\ and\ \citenamefont
  {Kitaev}(2005)}]{Bravyi05}%
  \BibitemOpen
  \bibfield  {author} {\bibinfo {author} {\bibfnamefont {S.}~\bibnamefont
  {Bravyi}}\ and\ \bibinfo {author} {\bibfnamefont {A.}~\bibnamefont
  {Kitaev}},\ }\href@noop {} {\bibfield  {journal} {\bibinfo  {journal} {Phys.
  Rev. A}\ }\textbf {\bibinfo {volume} {71}},\ \bibinfo {pages} {022316}
  (\bibinfo {year} {2005})}\BibitemShut {NoStop}%
\bibitem [{\citenamefont {Horsman}\ \emph {et~al.}(2012)\citenamefont
  {Horsman}, \citenamefont {Fowler}, \citenamefont {Devitt},\ and\
  \citenamefont {Meter}}]{Horsman12}%
  \BibitemOpen
  \bibfield  {author} {\bibinfo {author} {\bibfnamefont {C.}~\bibnamefont
  {Horsman}}, \bibinfo {author} {\bibfnamefont {A.~G.}\ \bibnamefont {Fowler}},
  \bibinfo {author} {\bibfnamefont {S.}~\bibnamefont {Devitt}}, \ and\ \bibinfo
  {author} {\bibfnamefont {R.~V.}\ \bibnamefont {Meter}},\ }\href@noop {}
  {\bibfield  {journal} {\bibinfo  {journal} {New J. Phys.}\ }\textbf {\bibinfo
  {volume} {14}},\ \bibinfo {pages} {123011} (\bibinfo {year}
  {2012})}\BibitemShut {NoStop}%
\bibitem [{\citenamefont {Brown}\ \emph {et~al.}(2017)\citenamefont {Brown},
  \citenamefont {Laubscher}, \citenamefont {Kesselring},\ and\ \citenamefont
  {Wootton}}]{Brown17}%
  \BibitemOpen
  \bibfield  {author} {\bibinfo {author} {\bibfnamefont {B.~J.}\ \bibnamefont
  {Brown}}, \bibinfo {author} {\bibfnamefont {K.}~\bibnamefont {Laubscher}},
  \bibinfo {author} {\bibfnamefont {M.~S.}\ \bibnamefont {Kesselring}}, \ and\
  \bibinfo {author} {\bibfnamefont {J.~R.}\ \bibnamefont {Wootton}},\
  }\href@noop {} {\bibfield  {journal} {\bibinfo  {journal} {Phys. Rev. X}\
  }\textbf {\bibinfo {volume} {7}},\ \bibinfo {pages} {021029} (\bibinfo {year}
  {2017})}\BibitemShut {NoStop}%
\bibitem [{\citenamefont {Brown}(2019)}]{Brown19a}%
  \BibitemOpen
  \bibfield  {author} {\bibinfo {author} {\bibfnamefont {B.~J.}\ \bibnamefont
  {Brown}},\ }\href@noop {} {\bibfield  {journal} {\bibinfo  {journal}
  {arXiv:1903.11634}\ } (\bibinfo {year} {2019})}\BibitemShut {NoStop}%
\bibitem [{\citenamefont {Barends}\ \emph {et~al.}(2014)\citenamefont
  {Barends}, \citenamefont {Kelly}, \citenamefont {Megrant}, \citenamefont
  {Veitia}, \citenamefont {Sank}, \citenamefont {Jeffry}, \citenamefont
  {White}, \citenamefont {Mutus}, \citenamefont {Fowler}, \citenamefont
  {Campbell}, \citenamefont {Chen}, \citenamefont {Chen}, \citenamefont
  {Chiaro}, \citenamefont {Dunsworth}, \citenamefont {Neill}, \citenamefont
  {O'Malley}, \citenamefont {Roushan}, \citenamefont {Vainsencher},
  \citenamefont {Wenner}, \citenamefont {Korotkov}, \citenamefont {Cleland},\
  and\ \citenamefont {Martinis}}]{Barends14}%
  \BibitemOpen
  \bibfield  {author} {\bibinfo {author} {\bibfnamefont {R.}~\bibnamefont
  {Barends}}, \bibinfo {author} {\bibfnamefont {J.}~\bibnamefont {Kelly}},
  \bibinfo {author} {\bibfnamefont {A.}~\bibnamefont {Megrant}}, \bibinfo
  {author} {\bibfnamefont {A.}~\bibnamefont {Veitia}}, \bibinfo {author}
  {\bibfnamefont {D.}~\bibnamefont {Sank}}, \bibinfo {author} {\bibfnamefont
  {E.}~\bibnamefont {Jeffry}}, \bibinfo {author} {\bibfnamefont {T.~C.}\
  \bibnamefont {White}}, \bibinfo {author} {\bibfnamefont {J.}~\bibnamefont
  {Mutus}}, \bibinfo {author} {\bibfnamefont {A.~G.}\ \bibnamefont {Fowler}},
  \bibinfo {author} {\bibfnamefont {B.}~\bibnamefont {Campbell}}, \bibinfo
  {author} {\bibfnamefont {Y.}~\bibnamefont {Chen}}, \bibinfo {author}
  {\bibfnamefont {Z.}~\bibnamefont {Chen}}, \bibinfo {author} {\bibfnamefont
  {B.}~\bibnamefont {Chiaro}}, \bibinfo {author} {\bibfnamefont
  {A.}~\bibnamefont {Dunsworth}}, \bibinfo {author} {\bibfnamefont
  {C.}~\bibnamefont {Neill}}, \bibinfo {author} {\bibfnamefont
  {P.}~\bibnamefont {O'Malley}}, \bibinfo {author} {\bibfnamefont
  {P.}~\bibnamefont {Roushan}}, \bibinfo {author} {\bibfnamefont
  {A.}~\bibnamefont {Vainsencher}}, \bibinfo {author} {\bibfnamefont
  {J.}~\bibnamefont {Wenner}}, \bibinfo {author} {\bibfnamefont {A.~N.}\
  \bibnamefont {Korotkov}}, \bibinfo {author} {\bibfnamefont {A.~N.}\
  \bibnamefont {Cleland}}, \ and\ \bibinfo {author} {\bibfnamefont {J.~M.}\
  \bibnamefont {Martinis}},\ }\href@noop {} {\bibfield  {journal} {\bibinfo
  {journal} {Nature}\ }\textbf {\bibinfo {volume} {508}},\ \bibinfo {pages}
  {500} (\bibinfo {year} {2014})}\BibitemShut {NoStop}%
\bibitem [{\citenamefont {C{\'o}rcoles}\ \emph {et~al.}(2015)\citenamefont
  {C{\'o}rcoles}, \citenamefont {Magesan}, \citenamefont {Srinivasan},
  \citenamefont {Cross}, \citenamefont {Steffen}, \citenamefont {Gambetta},\
  and\ \citenamefont {Chow}}]{Corcoles15}%
  \BibitemOpen
  \bibfield  {author} {\bibinfo {author} {\bibfnamefont {A.~D.}\ \bibnamefont
  {C{\'o}rcoles}}, \bibinfo {author} {\bibfnamefont {E.}~\bibnamefont
  {Magesan}}, \bibinfo {author} {\bibfnamefont {S.~J.}\ \bibnamefont
  {Srinivasan}}, \bibinfo {author} {\bibfnamefont {A.~W.}\ \bibnamefont
  {Cross}}, \bibinfo {author} {\bibfnamefont {M.}~\bibnamefont {Steffen}},
  \bibinfo {author} {\bibfnamefont {J.~M.}\ \bibnamefont {Gambetta}}, \ and\
  \bibinfo {author} {\bibfnamefont {J.~M.}\ \bibnamefont {Chow}},\ }\href@noop
  {} {\bibfield  {journal} {\bibinfo  {journal} {Nat. Comms.}\ }\textbf
  {\bibinfo {volume} {6}},\ \bibinfo {pages} {6979} (\bibinfo {year}
  {2015})}\BibitemShut {NoStop}%
\bibitem [{\citenamefont {Kelly}\ \emph {et~al.}(2015)\citenamefont {Kelly},
  \citenamefont {Barrends}, \citenamefont {Fowler}, \citenamefont {Megrant},
  \citenamefont {Jeffrey}, \citenamefont {White}, \citenamefont {Sank},
  \citenamefont {Mutus}, \citenamefont {Campbell}, \citenamefont {Chen},
  \citenamefont {Chen}, \citenamefont {Chiaro}, \citenamefont {Dunsworth},
  \citenamefont {Hoi}, \citenamefont {Neill}, \citenamefont {O'Malley},
  \citenamefont {Quintana}, \citenamefont {Roushan}, \citenamefont {Wenner},
  \citenamefont {Cleland},\ and\ \citenamefont {Martinis}}]{Kelly15}%
  \BibitemOpen
  \bibfield  {author} {\bibinfo {author} {\bibfnamefont {J.}~\bibnamefont
  {Kelly}}, \bibinfo {author} {\bibfnamefont {R.}~\bibnamefont {Barrends}},
  \bibinfo {author} {\bibfnamefont {A.~G.}\ \bibnamefont {Fowler}}, \bibinfo
  {author} {\bibfnamefont {A.}~\bibnamefont {Megrant}}, \bibinfo {author}
  {\bibfnamefont {E.}~\bibnamefont {Jeffrey}}, \bibinfo {author} {\bibfnamefont
  {T.~C.}\ \bibnamefont {White}}, \bibinfo {author} {\bibfnamefont
  {D.}~\bibnamefont {Sank}}, \bibinfo {author} {\bibfnamefont {J.~Y.}\
  \bibnamefont {Mutus}}, \bibinfo {author} {\bibfnamefont {B.}~\bibnamefont
  {Campbell}}, \bibinfo {author} {\bibfnamefont {Y.}~\bibnamefont {Chen}},
  \bibinfo {author} {\bibfnamefont {Z.}~\bibnamefont {Chen}}, \bibinfo {author}
  {\bibfnamefont {B.}~\bibnamefont {Chiaro}}, \bibinfo {author} {\bibfnamefont
  {A.}~\bibnamefont {Dunsworth}}, \bibinfo {author} {\bibfnamefont {I.-C.}\
  \bibnamefont {Hoi}}, \bibinfo {author} {\bibfnamefont {C.}~\bibnamefont
  {Neill}}, \bibinfo {author} {\bibfnamefont {P.~J.~J.}\ \bibnamefont
  {O'Malley}}, \bibinfo {author} {\bibfnamefont {C.}~\bibnamefont {Quintana}},
  \bibinfo {author} {\bibfnamefont {P.}~\bibnamefont {Roushan}}, \bibinfo
  {author} {\bibfnamefont {A.~V.~J.}\ \bibnamefont {Wenner}}, \bibinfo {author}
  {\bibfnamefont {A.~N.}\ \bibnamefont {Cleland}}, \ and\ \bibinfo {author}
  {\bibfnamefont {J.~M.}\ \bibnamefont {Martinis}},\ }\href@noop {} {\bibfield
  {journal} {\bibinfo  {journal} {Nature}\ }\textbf {\bibinfo {volume} {519}},\
  \bibinfo {pages} {66} (\bibinfo {year} {2015})}\BibitemShut {NoStop}%
\bibitem [{\citenamefont {Takita}\ \emph {et~al.}(2016)\citenamefont {Takita},
  \citenamefont {C\'{o}rcoles}, \citenamefont {Magesan}, \citenamefont {Abdo},
  \citenamefont {Brink}, \citenamefont {Cross}, \citenamefont {Chow},\ and\
  \citenamefont {Gambetta}}]{Takita16}%
  \BibitemOpen
  \bibfield  {author} {\bibinfo {author} {\bibfnamefont {M.}~\bibnamefont
  {Takita}}, \bibinfo {author} {\bibfnamefont {A.~D.}\ \bibnamefont
  {C\'{o}rcoles}}, \bibinfo {author} {\bibfnamefont {E.}~\bibnamefont
  {Magesan}}, \bibinfo {author} {\bibfnamefont {B.}~\bibnamefont {Abdo}},
  \bibinfo {author} {\bibfnamefont {M.}~\bibnamefont {Brink}}, \bibinfo
  {author} {\bibfnamefont {A.~W.}\ \bibnamefont {Cross}}, \bibinfo {author}
  {\bibfnamefont {J.~M.}\ \bibnamefont {Chow}}, \ and\ \bibinfo {author}
  {\bibfnamefont {J.~M.}\ \bibnamefont {Gambetta}},\ }\href@noop {} {\bibfield
  {journal} {\bibinfo  {journal} {Phys. Rev. Lett.}\ }\textbf {\bibinfo
  {volume} {117}},\ \bibinfo {pages} {210505} (\bibinfo {year}
  {2016})}\BibitemShut {NoStop}%
\bibitem [{\citenamefont {Duclos-Cianci}\ and\ \citenamefont
  {Poulin}(2010)}]{DuclosCianci10}%
  \BibitemOpen
  \bibfield  {author} {\bibinfo {author} {\bibfnamefont {G.}~\bibnamefont
  {Duclos-Cianci}}\ and\ \bibinfo {author} {\bibfnamefont {D.}~\bibnamefont
  {Poulin}},\ }\href@noop {} {\bibfield  {journal} {\bibinfo  {journal} {Phys.
  Rev. Lett.}\ }\textbf {\bibinfo {volume} {104}},\ \bibinfo {pages} {050504}
  (\bibinfo {year} {2010})}\BibitemShut {NoStop}%
\bibitem [{\citenamefont {Fowler}\ \emph
  {et~al.}(2012{\natexlab{a}})\citenamefont {Fowler}, \citenamefont
  {Mariantoni}, \citenamefont {Martinis},\ and\ \citenamefont
  {Cleland}}]{Fowler12a}%
  \BibitemOpen
  \bibfield  {author} {\bibinfo {author} {\bibfnamefont {A.~G.}\ \bibnamefont
  {Fowler}}, \bibinfo {author} {\bibfnamefont {M.}~\bibnamefont {Mariantoni}},
  \bibinfo {author} {\bibfnamefont {J.~M.}\ \bibnamefont {Martinis}}, \ and\
  \bibinfo {author} {\bibfnamefont {A.~N.}\ \bibnamefont {Cleland}},\
  }\href@noop {} {\bibfield  {journal} {\bibinfo  {journal} {Phys. Rev. A}\
  }\textbf {\bibinfo {volume} {86}},\ \bibinfo {pages} {032324} (\bibinfo
  {year} {2012}{\natexlab{a}})}\BibitemShut {NoStop}%
\bibitem [{\citenamefont {Bravyi}\ \emph {et~al.}(2014)\citenamefont {Bravyi},
  \citenamefont {Suchara},\ and\ \citenamefont {Vargo}}]{Bravyi14}%
  \BibitemOpen
  \bibfield  {author} {\bibinfo {author} {\bibfnamefont {S.}~\bibnamefont
  {Bravyi}}, \bibinfo {author} {\bibfnamefont {M.}~\bibnamefont {Suchara}}, \
  and\ \bibinfo {author} {\bibfnamefont {A.}~\bibnamefont {Vargo}},\
  }\href@noop {} {\bibfield  {journal} {\bibinfo  {journal} {Phys. Rev. A}\
  }\textbf {\bibinfo {volume} {90}},\ \bibinfo {pages} {032326} (\bibinfo
  {year} {2014})}\BibitemShut {NoStop}%
\bibitem [{\citenamefont {Darmawan}\ and\ \citenamefont
  {Poulin}(2018)}]{Darmawan18}%
  \BibitemOpen
  \bibfield  {author} {\bibinfo {author} {\bibfnamefont {A.~S.}\ \bibnamefont
  {Darmawan}}\ and\ \bibinfo {author} {\bibfnamefont {D.}~\bibnamefont
  {Poulin}},\ }\href@noop {} {\bibfield  {journal} {\bibinfo  {journal} {Phys.
  Rev. E}\ }\textbf {\bibinfo {volume} {97}},\ \bibinfo {pages} {051302}
  (\bibinfo {year} {2018})}\BibitemShut {NoStop}%
\bibitem [{\citenamefont {Edmonds}(1965)}]{Edmonds65}%
  \BibitemOpen
  \bibfield  {author} {\bibinfo {author} {\bibfnamefont {J.}~\bibnamefont
  {Edmonds}},\ }\href@noop {} {\bibfield  {journal} {\bibinfo  {journal}
  {Canad. J. Math.}\ }\textbf {\bibinfo {volume} {17}},\ \bibinfo {pages} {449}
  (\bibinfo {year} {1965})}\BibitemShut {NoStop}%
\bibitem [{\citenamefont {Kolmogorov}(2009)}]{Kolmogorov09}%
  \BibitemOpen
  \bibfield  {author} {\bibinfo {author} {\bibfnamefont {V.}~\bibnamefont
  {Kolmogorov}},\ }\href@noop {} {\bibfield  {journal} {\bibinfo  {journal}
  {Math. Prog. Comp.}\ }\textbf {\bibinfo {volume} {1}},\ \bibinfo {pages} {43}
  (\bibinfo {year} {2009})}\BibitemShut {NoStop}%
\bibitem [{\citenamefont {Ohno}\ \emph {et~al.}(2004)\citenamefont {Ohno},
  \citenamefont {Arakawa}, \citenamefont {Ichinose},\ and\ \citenamefont
  {Matsui}}]{Ohno04}%
  \BibitemOpen
  \bibfield  {author} {\bibinfo {author} {\bibfnamefont {T.}~\bibnamefont
  {Ohno}}, \bibinfo {author} {\bibfnamefont {G.}~\bibnamefont {Arakawa}},
  \bibinfo {author} {\bibfnamefont {I.}~\bibnamefont {Ichinose}}, \ and\
  \bibinfo {author} {\bibfnamefont {T.}~\bibnamefont {Matsui}},\ }\href@noop {}
  {\bibfield  {journal} {\bibinfo  {journal} {Nucl. Phys. B}\ }\textbf
  {\bibinfo {volume} {697}},\ \bibinfo {pages} {462} (\bibinfo {year}
  {2004})}\BibitemShut {NoStop}%
\bibitem [{\citenamefont {Kubica}\ \emph {et~al.}(2018)\citenamefont {Kubica},
  \citenamefont {Beverland}, \citenamefont {Brand{\~a}o}, \citenamefont
  {Preskill},\ and\ \citenamefont {Svore}}]{Kubica17}%
  \BibitemOpen
  \bibfield  {author} {\bibinfo {author} {\bibfnamefont {A.}~\bibnamefont
  {Kubica}}, \bibinfo {author} {\bibfnamefont {M.~E.}\ \bibnamefont
  {Beverland}}, \bibinfo {author} {\bibfnamefont {F.}~\bibnamefont
  {Brand{\~a}o}}, \bibinfo {author} {\bibfnamefont {J.}~\bibnamefont
  {Preskill}}, \ and\ \bibinfo {author} {\bibfnamefont {K.~M.}\ \bibnamefont
  {Svore}},\ }\href@noop {} {\bibfield  {journal} {\bibinfo  {journal} {Phys.
  Rev. Lett.}\ }\textbf {\bibinfo {volume} {120}},\ \bibinfo {pages} {180501}
  (\bibinfo {year} {2018})}\BibitemShut {NoStop}%
\bibitem [{\citenamefont {Wootton}\ and\ \citenamefont
  {Loss}(2012)}]{Wootton12}%
  \BibitemOpen
  \bibfield  {author} {\bibinfo {author} {\bibfnamefont {J.~R.}\ \bibnamefont
  {Wootton}}\ and\ \bibinfo {author} {\bibfnamefont {D.}~\bibnamefont {Loss}},\
  }\href@noop {} {\bibfield  {journal} {\bibinfo  {journal} {Phys. Rev. Lett.}\
  }\textbf {\bibinfo {volume} {109}},\ \bibinfo {pages} {160503} (\bibinfo
  {year} {2012})}\BibitemShut {NoStop}%
\bibitem [{\citenamefont {Fowler}(2013)}]{Fowler13}%
  \BibitemOpen
  \bibfield  {author} {\bibinfo {author} {\bibfnamefont {A.~G.}\ \bibnamefont
  {Fowler}},\ }\href@noop {} {\bibfield  {journal} {\bibinfo  {journal}
  {arXiv:1310.0863}\ } (\bibinfo {year} {2013})}\BibitemShut {NoStop}%
\bibitem [{\citenamefont {Hutter}\ and\ \citenamefont
  {Loss}(2014)}]{Hutter14a}%
  \BibitemOpen
  \bibfield  {author} {\bibinfo {author} {\bibfnamefont {A.}~\bibnamefont
  {Hutter}}\ and\ \bibinfo {author} {\bibfnamefont {D.}~\bibnamefont {Loss}},\
  }\href@noop {} {\bibfield  {journal} {\bibinfo  {journal} {Phys. Rev. A}\
  }\textbf {\bibinfo {volume} {89}},\ \bibinfo {pages} {042334} (\bibinfo
  {year} {2014})}\BibitemShut {NoStop}%
\bibitem [{\citenamefont {Nickerson}\ and\ \citenamefont
  {Brown}(2019)}]{Nickerson17}%
  \BibitemOpen
  \bibfield  {author} {\bibinfo {author} {\bibfnamefont {N.~H.}\ \bibnamefont
  {Nickerson}}\ and\ \bibinfo {author} {\bibfnamefont {B.~J.}\ \bibnamefont
  {Brown}},\ }\href@noop {} {\bibfield  {journal} {\bibinfo  {journal}
  {Quantum}\ }\textbf {\bibinfo {volume} {3}},\ \bibinfo {pages} {131}
  (\bibinfo {year} {2019})}\BibitemShut {NoStop}%
\bibitem [{\citenamefont {Fowler}\ \emph
  {et~al.}(2012{\natexlab{b}})\citenamefont {Fowler}, \citenamefont
  {Whiteside}, \citenamefont {McInnes},\ and\ \citenamefont
  {Rabbani}}]{Fowler12b}%
  \BibitemOpen
  \bibfield  {author} {\bibinfo {author} {\bibfnamefont {A.~G.}\ \bibnamefont
  {Fowler}}, \bibinfo {author} {\bibfnamefont {A.~C.}\ \bibnamefont
  {Whiteside}}, \bibinfo {author} {\bibfnamefont {A.~L.}\ \bibnamefont
  {McInnes}}, \ and\ \bibinfo {author} {\bibfnamefont {A.}~\bibnamefont
  {Rabbani}},\ }\href@noop {} {\bibfield  {journal} {\bibinfo  {journal} {Phys.
  Rev. X}\ }\textbf {\bibinfo {volume} {2}},\ \bibinfo {pages} {041003}
  (\bibinfo {year} {2012}{\natexlab{b}})}\BibitemShut {NoStop}%
\bibitem [{\citenamefont {Tuckett}\ \emph
  {et~al.}(2018{\natexlab{a}})\citenamefont {Tuckett}, \citenamefont
  {Bartlett},\ and\ \citenamefont {Flammia}}]{Tuckett18}%
  \BibitemOpen
  \bibfield  {author} {\bibinfo {author} {\bibfnamefont {D.~K.}\ \bibnamefont
  {Tuckett}}, \bibinfo {author} {\bibfnamefont {S.~D.}\ \bibnamefont
  {Bartlett}}, \ and\ \bibinfo {author} {\bibfnamefont {S.~T.}\ \bibnamefont
  {Flammia}},\ }\href@noop {} {\bibfield  {journal} {\bibinfo  {journal} {Phys.
  Rev. Lett.}\ }\textbf {\bibinfo {volume} {120}},\ \bibinfo {pages} {050505}
  (\bibinfo {year} {2018}{\natexlab{a}})}\BibitemShut {NoStop}%
\bibitem [{\citenamefont {Tuckett}\ \emph
  {et~al.}(2018{\natexlab{b}})\citenamefont {Tuckett}, \citenamefont {Chubb},
  \citenamefont {Bravyi}, \citenamefont {Bartlett},\ and\ \citenamefont
  {Flammia}}]{Tuckett18a}%
  \BibitemOpen
  \bibfield  {author} {\bibinfo {author} {\bibfnamefont {D.~K.}\ \bibnamefont
  {Tuckett}}, \bibinfo {author} {\bibfnamefont {C.~T.}\ \bibnamefont {Chubb}},
  \bibinfo {author} {\bibfnamefont {S.}~\bibnamefont {Bravyi}}, \bibinfo
  {author} {\bibfnamefont {S.~D.}\ \bibnamefont {Bartlett}}, \ and\ \bibinfo
  {author} {\bibfnamefont {S.~T.}\ \bibnamefont {Flammia}},\ }\href@noop {}
  {\bibfield  {journal} {\bibinfo  {journal} {arXiv:1812.08186}\ } (\bibinfo
  {year} {2018}{\natexlab{b}})}\BibitemShut {NoStop}%
\bibitem [{\citenamefont {Brown}\ and\ \citenamefont
  {Williamson}(2020)}]{Brown19}%
  \BibitemOpen
  \bibfield  {author} {\bibinfo {author} {\bibfnamefont {B.~J.}\ \bibnamefont
  {Brown}}\ and\ \bibinfo {author} {\bibfnamefont {D.~J.}\ \bibnamefont
  {Williamson}},\ }\href@noop {} {\bibfield  {journal} {\bibinfo  {journal}
  {Phys. Rev. Research}\ }\textbf {\bibinfo {volume} {2}},\ \bibinfo {pages}
  {013303} (\bibinfo {year} {2020})}\BibitemShut {NoStop}%
\bibitem [{\citenamefont {Delfosse}\ and\ \citenamefont
  {Nickerson}(2017)}]{Delfosse17}%
  \BibitemOpen
  \bibfield  {author} {\bibinfo {author} {\bibfnamefont {N.}~\bibnamefont
  {Delfosse}}\ and\ \bibinfo {author} {\bibfnamefont {N.~H.}\ \bibnamefont
  {Nickerson}},\ }\href@noop {} {\bibfield  {journal} {\bibinfo  {journal}
  {arXiv:1709.06218}\ } (\bibinfo {year} {2017})}\BibitemShut {NoStop}%
\bibitem [{\citenamefont {Raussendorf}\ \emph {et~al.}(2006)\citenamefont
  {Raussendorf}, \citenamefont {Harrington},\ and\ \citenamefont
  {Goyal}}]{Raussendorf06}%
  \BibitemOpen
  \bibfield  {author} {\bibinfo {author} {\bibfnamefont {R.}~\bibnamefont
  {Raussendorf}}, \bibinfo {author} {\bibfnamefont {J.}~\bibnamefont
  {Harrington}}, \ and\ \bibinfo {author} {\bibfnamefont {K.}~\bibnamefont
  {Goyal}},\ }\href@noop {} {\bibfield  {journal} {\bibinfo  {journal} {Ann.
  Phys.}\ }\textbf {\bibinfo {volume} {321}},\ \bibinfo {pages} {2242}
  (\bibinfo {year} {2006})}\BibitemShut {NoStop}%
\bibitem [{\citenamefont {Vuillot}\ \emph {et~al.}(2019)\citenamefont
  {Vuillot}, \citenamefont {Lao}, \citenamefont {Criger}, \citenamefont
  {Almud\'{e}ver}, \citenamefont {Bertels},\ and\ \citenamefont
  {Terhal}}]{Vuillot18}%
  \BibitemOpen
  \bibfield  {author} {\bibinfo {author} {\bibfnamefont {C.}~\bibnamefont
  {Vuillot}}, \bibinfo {author} {\bibfnamefont {L.}~\bibnamefont {Lao}},
  \bibinfo {author} {\bibfnamefont {B.}~\bibnamefont {Criger}}, \bibinfo
  {author} {\bibfnamefont {C.~G.}\ \bibnamefont {Almud\'{e}ver}}, \bibinfo
  {author} {\bibfnamefont {K.}~\bibnamefont {Bertels}}, \ and\ \bibinfo
  {author} {\bibfnamefont {B.~M.}\ \bibnamefont {Terhal}},\ }\href@noop {}
  {\bibfield  {journal} {\bibinfo  {journal} {New J. Phys.}\ }\textbf {\bibinfo
  {volume} {21}},\ \bibinfo {pages} {033028} (\bibinfo {year}
  {2019})}\BibitemShut {NoStop}%
\bibitem [{\citenamefont {Aliferis}\ and\ \citenamefont
  {Preskill}(2008)}]{Aliferis08}%
  \BibitemOpen
  \bibfield  {author} {\bibinfo {author} {\bibfnamefont {P.}~\bibnamefont
  {Aliferis}}\ and\ \bibinfo {author} {\bibfnamefont {J.}~\bibnamefont
  {Preskill}},\ }\href@noop {} {\bibfield  {journal} {\bibinfo  {journal}
  {Phys. Rev. A}\ }\textbf {\bibinfo {volume} {78}},\ \bibinfo {pages} {052331}
  (\bibinfo {year} {2008})}\BibitemShut {NoStop}%
\bibitem [{\citenamefont {Puri}\ \emph {et~al.}(2019)\citenamefont {Puri},
  \citenamefont {St-Jean}, \citenamefont {Gross}, \citenamefont {Grimm},
  \citenamefont {Frattini}, \citenamefont {Iyer}, \citenamefont {Krishna},
  \citenamefont {Touzard}, \citenamefont {Jiang}, \citenamefont {Blais},
  \citenamefont {Flammia},\ and\ \citenamefont {Girvin}}]{Puri19}%
  \BibitemOpen
  \bibfield  {author} {\bibinfo {author} {\bibfnamefont {S.}~\bibnamefont
  {Puri}}, \bibinfo {author} {\bibfnamefont {L.}~\bibnamefont {St-Jean}},
  \bibinfo {author} {\bibfnamefont {J.~A.}\ \bibnamefont {Gross}}, \bibinfo
  {author} {\bibfnamefont {A.}~\bibnamefont {Grimm}}, \bibinfo {author}
  {\bibfnamefont {N.~E.}\ \bibnamefont {Frattini}}, \bibinfo {author}
  {\bibfnamefont {P.~S.}\ \bibnamefont {Iyer}}, \bibinfo {author}
  {\bibfnamefont {A.}~\bibnamefont {Krishna}}, \bibinfo {author} {\bibfnamefont
  {S.}~\bibnamefont {Touzard}}, \bibinfo {author} {\bibfnamefont
  {L.}~\bibnamefont {Jiang}}, \bibinfo {author} {\bibfnamefont
  {A.}~\bibnamefont {Blais}}, \bibinfo {author} {\bibfnamefont {S.~T.}\
  \bibnamefont {Flammia}}, \ and\ \bibinfo {author} {\bibfnamefont {S.~M.}\
  \bibnamefont {Girvin}},\ }\href@noop {} {\bibfield  {journal} {\bibinfo
  {journal} {arXiv:1905.00450}\ } (\bibinfo {year} {2019})}\BibitemShut
  {NoStop}%
\bibitem [{\citenamefont {Chubb}\ and\ \citenamefont
  {Flammia}(2019)}]{Chubb18}%
  \BibitemOpen
  \bibfield  {author} {\bibinfo {author} {\bibfnamefont {C.~T.}\ \bibnamefont
  {Chubb}}\ and\ \bibinfo {author} {\bibfnamefont {S.~T.}\ \bibnamefont
  {Flammia}},\ }\href@noop {} {\bibfield  {journal} {\bibinfo  {journal}
  {arXiv:1809.10704}\ } (\bibinfo {year} {2019})}\BibitemShut {NoStop}%
\bibitem [{\citenamefont {Beverland}\ \emph {et~al.}(2019)\citenamefont
  {Beverland}, \citenamefont {Brown}, \citenamefont {Kastoryano},\ and\
  \citenamefont {Marolleau}}]{Beverland18}%
  \BibitemOpen
  \bibfield  {author} {\bibinfo {author} {\bibfnamefont {M.~E.}\ \bibnamefont
  {Beverland}}, \bibinfo {author} {\bibfnamefont {B.~J.}\ \bibnamefont
  {Brown}}, \bibinfo {author} {\bibfnamefont {M.~J.}\ \bibnamefont
  {Kastoryano}}, \ and\ \bibinfo {author} {\bibfnamefont {Q.}~\bibnamefont
  {Marolleau}},\ }\href@noop {} {\bibfield  {journal} {\bibinfo  {journal} {J.
  Stat. Mech.:Theor. Exp.}\ }\textbf {\bibinfo {volume} {2019}},\ \bibinfo
  {pages} {073404} (\bibinfo {year} {2019})}\BibitemShut {NoStop}%
\bibitem [{\citenamefont {Stephens}\ \emph {et~al.}(2013)\citenamefont
  {Stephens}, \citenamefont {Munro},\ and\ \citenamefont
  {Nemoto}}]{Stephens13}%
  \BibitemOpen
  \bibfield  {author} {\bibinfo {author} {\bibfnamefont {A.~M.}\ \bibnamefont
  {Stephens}}, \bibinfo {author} {\bibfnamefont {W.~J.}\ \bibnamefont {Munro}},
  \ and\ \bibinfo {author} {\bibfnamefont {K.}~\bibnamefont {Nemoto}},\
  }\href@noop {} {\bibfield  {journal} {\bibinfo  {journal} {Phys. Rev. A}\
  }\textbf {\bibinfo {volume} {88}},\ \bibinfo {pages} {060301(R)} (\bibinfo
  {year} {2013})}\BibitemShut {NoStop}%
\bibitem [{\citenamefont {Xu}\ \emph {et~al.}(2018)\citenamefont {Xu},
  \citenamefont {Ahao}, \citenamefont {Yuan},\ and\ \citenamefont
  {Benjamin}}]{Xu18}%
  \BibitemOpen
  \bibfield  {author} {\bibinfo {author} {\bibfnamefont {X.}~\bibnamefont
  {Xu}}, \bibinfo {author} {\bibfnamefont {Q.}~\bibnamefont {Ahao}}, \bibinfo
  {author} {\bibfnamefont {X.}~\bibnamefont {Yuan}}, \ and\ \bibinfo {author}
  {\bibfnamefont {S.~C.}\ \bibnamefont {Benjamin}},\ }\href@noop {} {\bibfield
  {journal} {\bibinfo  {journal} {arXiv:1812.01505}\ } (\bibinfo {year}
  {2018})}\BibitemShut {NoStop}%
\bibitem [{\citenamefont {Delfosse}(2014)}]{Delfosse14}%
  \BibitemOpen
  \bibfield  {author} {\bibinfo {author} {\bibfnamefont {N.}~\bibnamefont
  {Delfosse}},\ }\href@noop {} {\bibfield  {journal} {\bibinfo  {journal}
  {Phys. Rev. A}\ }\textbf {\bibinfo {volume} {89}},\ \bibinfo {pages} {012317}
  (\bibinfo {year} {2014})}\BibitemShut {NoStop}%
\bibitem [{\citenamefont {Kubica}\ and\ \citenamefont
  {Delfosse}(2019)}]{Kubica19}%
  \BibitemOpen
  \bibfield  {author} {\bibinfo {author} {\bibfnamefont {A.}~\bibnamefont
  {Kubica}}\ and\ \bibinfo {author} {\bibfnamefont {N.}~\bibnamefont
  {Delfosse}},\ }\href@noop {} {\bibfield  {journal} {\bibinfo  {journal}
  {arXiv:1905.07393}\ } (\bibinfo {year} {2019})}\BibitemShut {NoStop}%
\end{thebibliography}
%

\newpage

\section{Supplemental material}

Here, we provide technical details of the decoding algorithm implementation and some examples of its operation. This complements the conceptual description we give in the letter.

\subsection{Decoder implementation}
\label{f-sec:decoder-implementation}

Consider the tailored surface codes defined in the letter, with qubits on the faces and $X$- and $Y$-type stabilizers on the black and white vertices, respectively.
The input to the decoding algorithm is an error syndrome, and some assumed noise parameters.
The output of the decoding algorithm is a recovery operator that returns the code to the codespace.
The syndrome is generated from repeated stabilizer measurements~\cite{Dennis02}, such that \emph{defects} are identified with vertex locations in time where the parity of successive pairs of stabilizer measurements is odd.
We label defects as $X$- or $Y$-type depending on whether they are due to $X$- or $Y$-type stabilizer checks.

The decoding algorithm, see Algorithm~\ref{f-algo:decode}, exploits symmetries of the code and noise model in the following way.
A graph is constructed with two nodes, labeled horizontal (H) and vertical (V), for each defect.
Edges are added between similarly oriented nodes and weighted by a distance function, described in more detail below.
Nodes of the graph are paired using minimum-weight perfect matching (MWPM) and the defects identified with the nodes are grouped into \emph{clusters} by following the path traced out by matched pairs of nodes.
In the case of pure Z noise, these clusters are guaranteed to have an even parity of each type of defect; such even-parity clusters are locally correctable and we refer to them as \emph{neutral}.
In the case of finite bias, clusters are guaranteed to have the same parity of each type of defect but this parity may be odd; such odd-parity clusters are not locally correctable and we refer to them as \emph{charged}.
Having locally corrected the clusters as far as possible, the decoding algorithm invokes a residual decoding sub-algorithm, see Algorithm~\ref{f-algo:residual-decode}, that again uses MWPM to pair charged clusters, possibly through neutral clusters, ensuring the code is returned to the codespace.

\begin{figure*}[tb]
  \centering
  \hspace{.5cm}
  \subfloat{\includegraphics[width=0.228\textwidth]{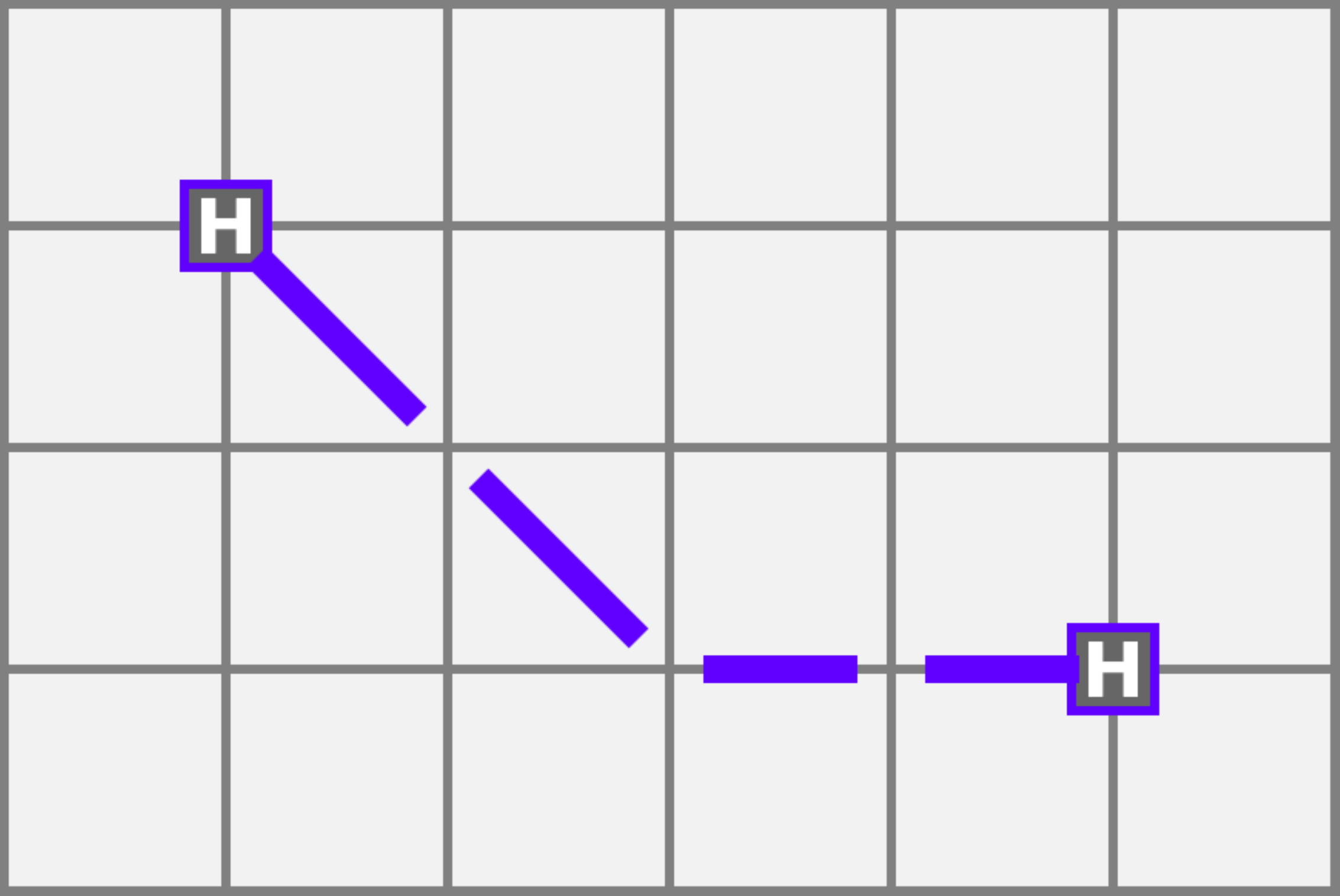}}
  \hfill
  \subfloat{\includegraphics[width=0.152\textwidth]{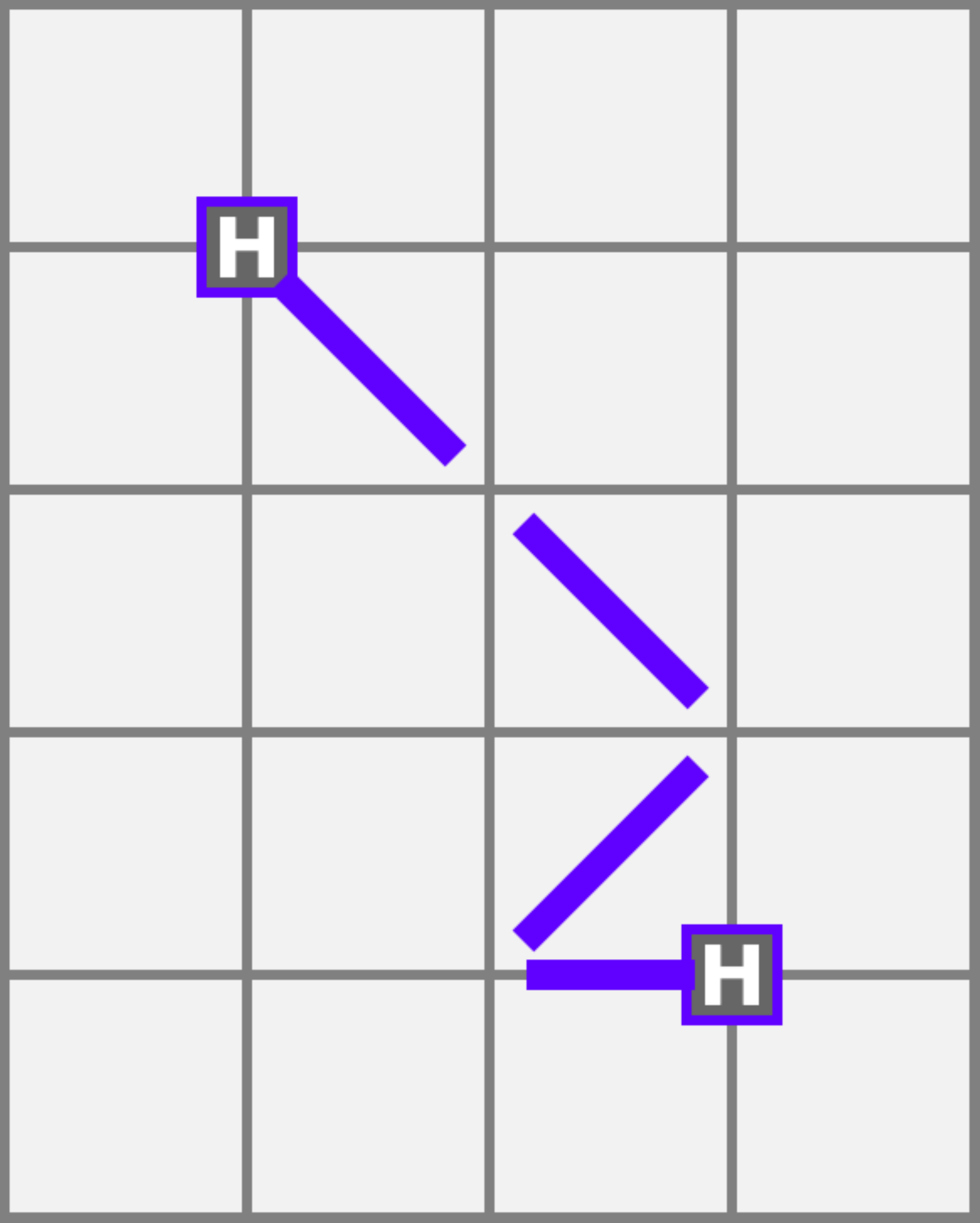}}
  \hfill
  \subfloat{\includegraphics[width=0.228\textwidth]{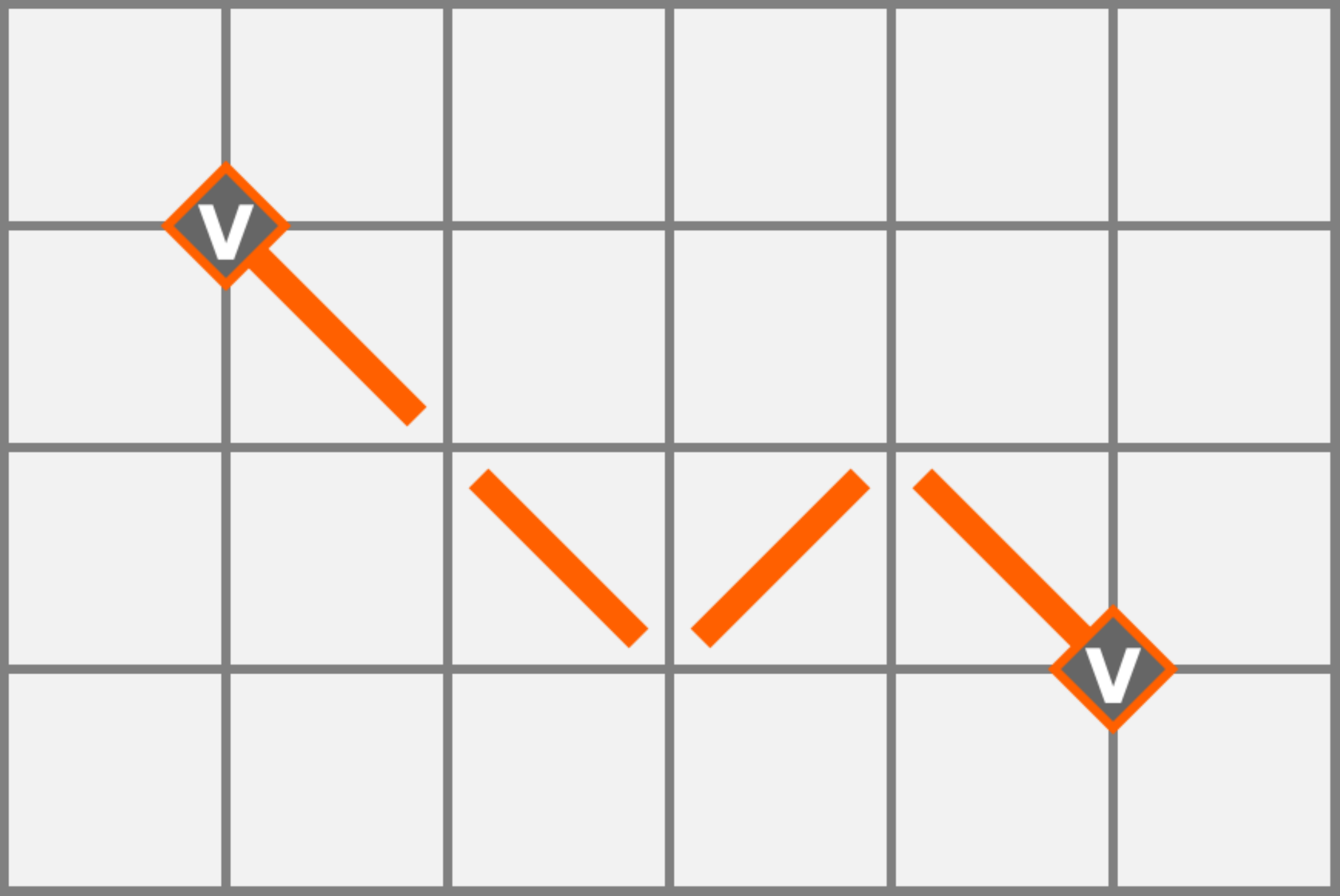}}
  \hfill
  \subfloat{\includegraphics[width=0.152\textwidth]{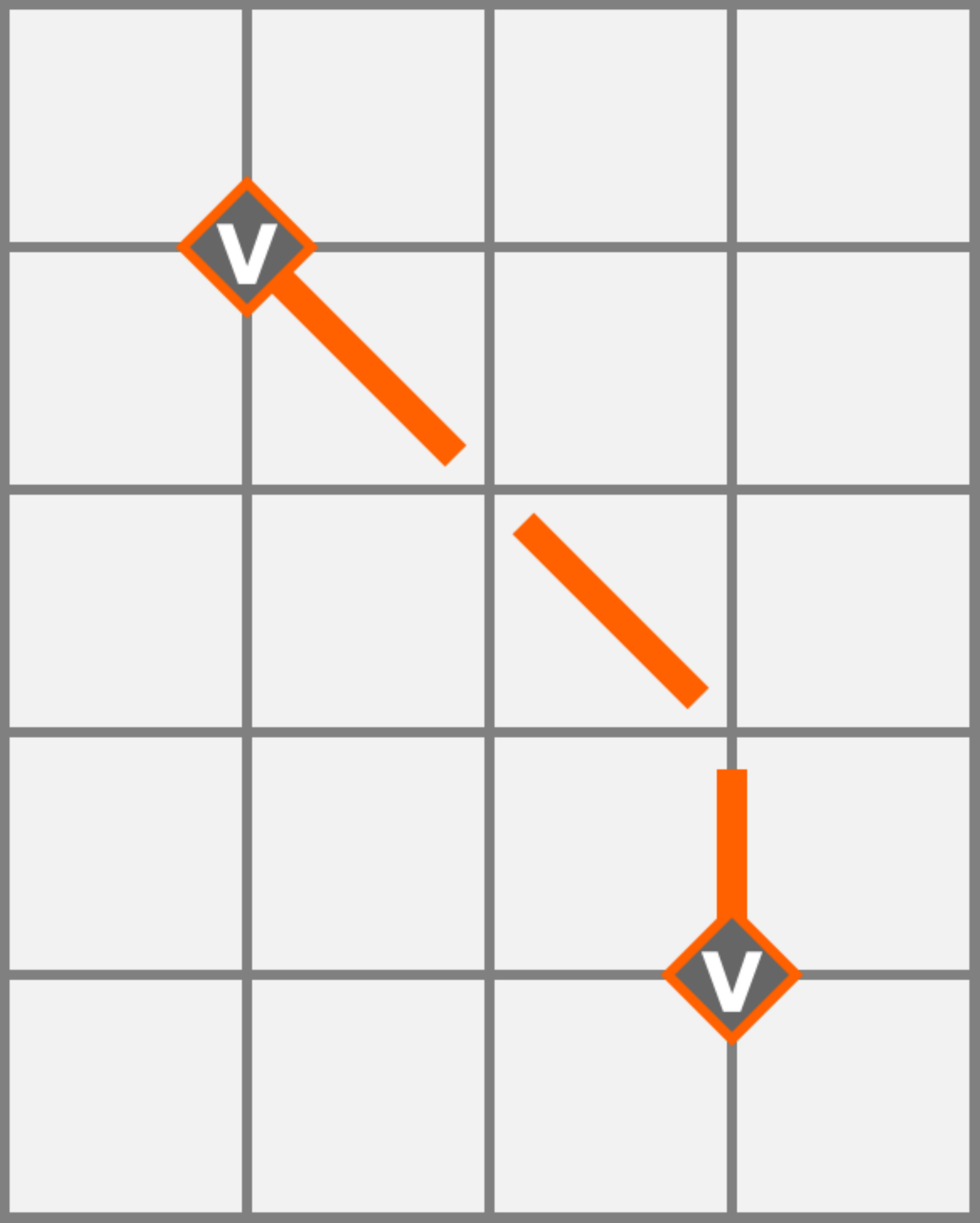}}
  \hspace{.5cm}
  \caption{Examples of paths between H and V nodes broken into parallel and diagonal steps, minimizing first the number of diagonal steps and then parallel steps; between H (V) nodes parallel steps are horizontal (vertical).
  }
  \label{f-fig:h-v-steps}
\end{figure*}

The graph used in the main decoding algorithm is key to exploiting the symmetries of the code and noise model.
As mentioned above, the graph contains a pair of nodes, labeled H and V, for each syndrome defect, and edges are added between similarly oriented nodes (i.e.\ H to H and V to V), with edge weights given by a distance function.
(For a code with boundaries, an additional pair of virtual H and V nodes is added and connected with zero weight, at each boundary vertex where a stabilizer is not applied.)
The distance function, see Algorithm~\ref{f-algo:distance}, breaks the path between nodes into time, parallel and diagonal steps.
Parallel steps are along horizontal (vertical) lines on the lattice between H (V) nodes.
Measurement errors displace defects in time steps, high-rate $Z$ errors displace defects in parallel steps, and low-rate $X$ or $Y$ errors displace defects in diagonal steps.
Each of these steps is weighted as a function of the noise parameters: bias, qubit error probability and measurement error probability.
The minimum distance between nodes is found by minimizing the number of diagonal, parallel and time steps in that order of priority, and then returning the weighted sum over these steps.
Figure~\ref{f-fig:h-v-steps} gives some examples of paths between H and V nodes broken into parallel and diagonal steps.

The derivation of step weights as functions of the noise parameters is as follows.
Consider the code, with stabilizer measurements repeated over a fixed number of time steps, represented by a (2+1)-dimensional lattice with a total of $N$ qubit locations (one for each physical qubit at each time step) and $M$ stabilizer measurement locations (one for each stabilizer measurement at each time step).
The probability of an error configuration, $E$, consisting of $H$ high-rate qubit errors, $L$ low-rate qubit errors and $Q$ measurement errors, is given by
\[
\Pr(E) = (1-p)^{N-H-L} p_\text{h.r.}^H p_\text{l.r.}^L (1-q)^{M-Q} q^Q
\]
where $p$, $p_\text{h.r.}$, $p_\text{l.r.}$ and $q$ are the probabilities of any qubit error, a high-rate qubit error, a low-rate qubit error, and a measurement error, respectively.
Noting that $N$ and $M$ are constant for a fixed number of time steps, we have
\[
\log \Pr(E) = H\log\frac{p_\text{h.r.}}{1-p} + L\log\frac{p_\text{l.r.}}{1-p} + Q\log\frac{q}{1-q} + k,
\]
where $k$ is a constant independent of the error configuration.
Recalling, from the letter, that $p_\text{h.r.}=p\eta/(\eta +1)$ and $p_\text{l.r.}=p/(2(\eta+1))$, where $\eta$ is the noise bias, we see that the time, parallel and diagonal step weights, as functions of $\eta$, $p$ and $q$ can be defined, respectively, as
\begin{align*}
\mu_t &= -\log[q/(1-q)] \\
\mu_p &= -\log[\eta/(\eta+1)]-\log[p/(1-p)] \\
\mu_d &= -\log[1/(2(\eta+1))]-\log[p/(1-p)].
\end{align*}

\begin{figure*}[tb]
  \centering
  \subfloat[]{\includegraphics[width=0.19\textwidth]{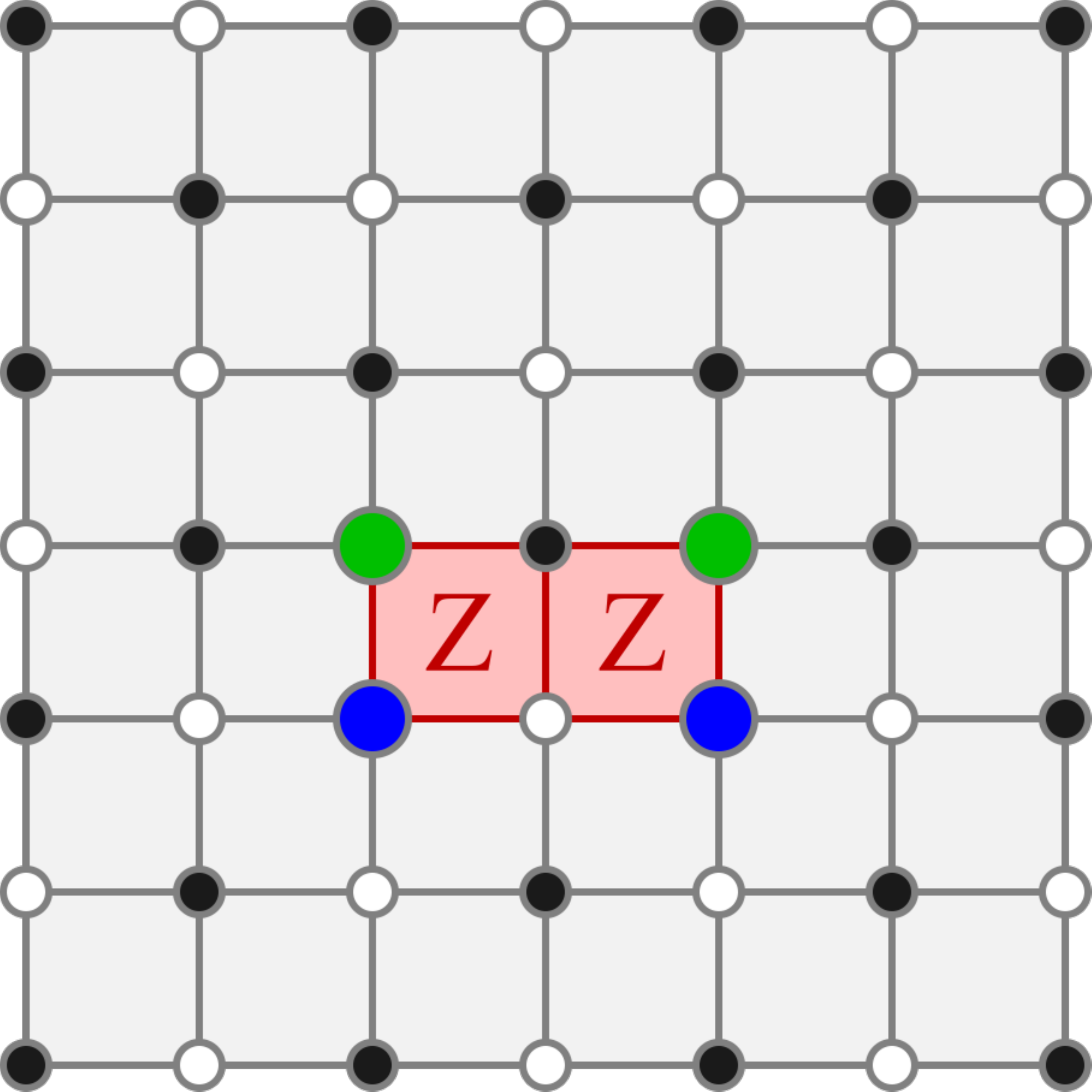}}
  \hfill
  \subfloat[]{\includegraphics[width=0.19\textwidth]{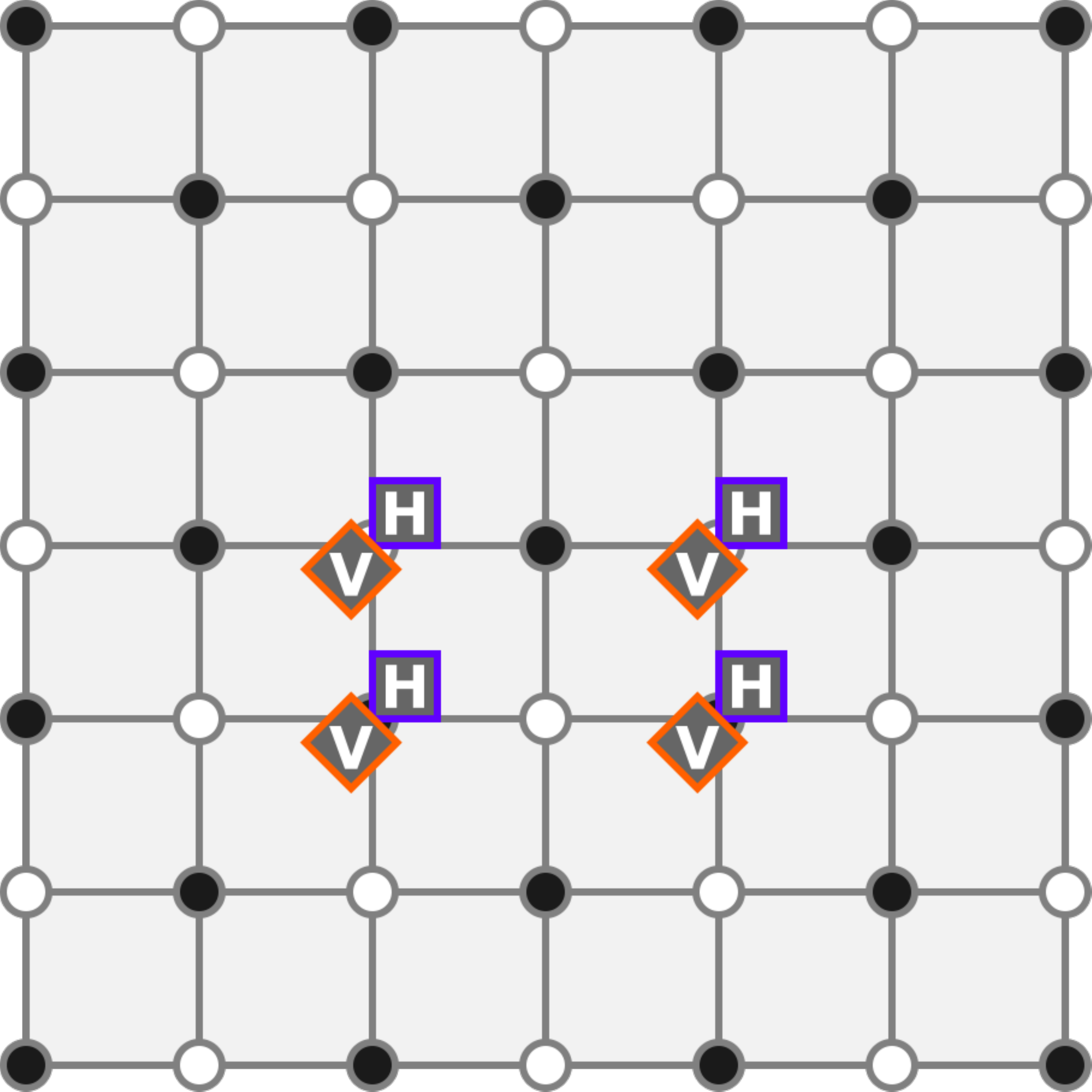}}
  \hfill
  \subfloat[]{\includegraphics[width=0.19\textwidth]{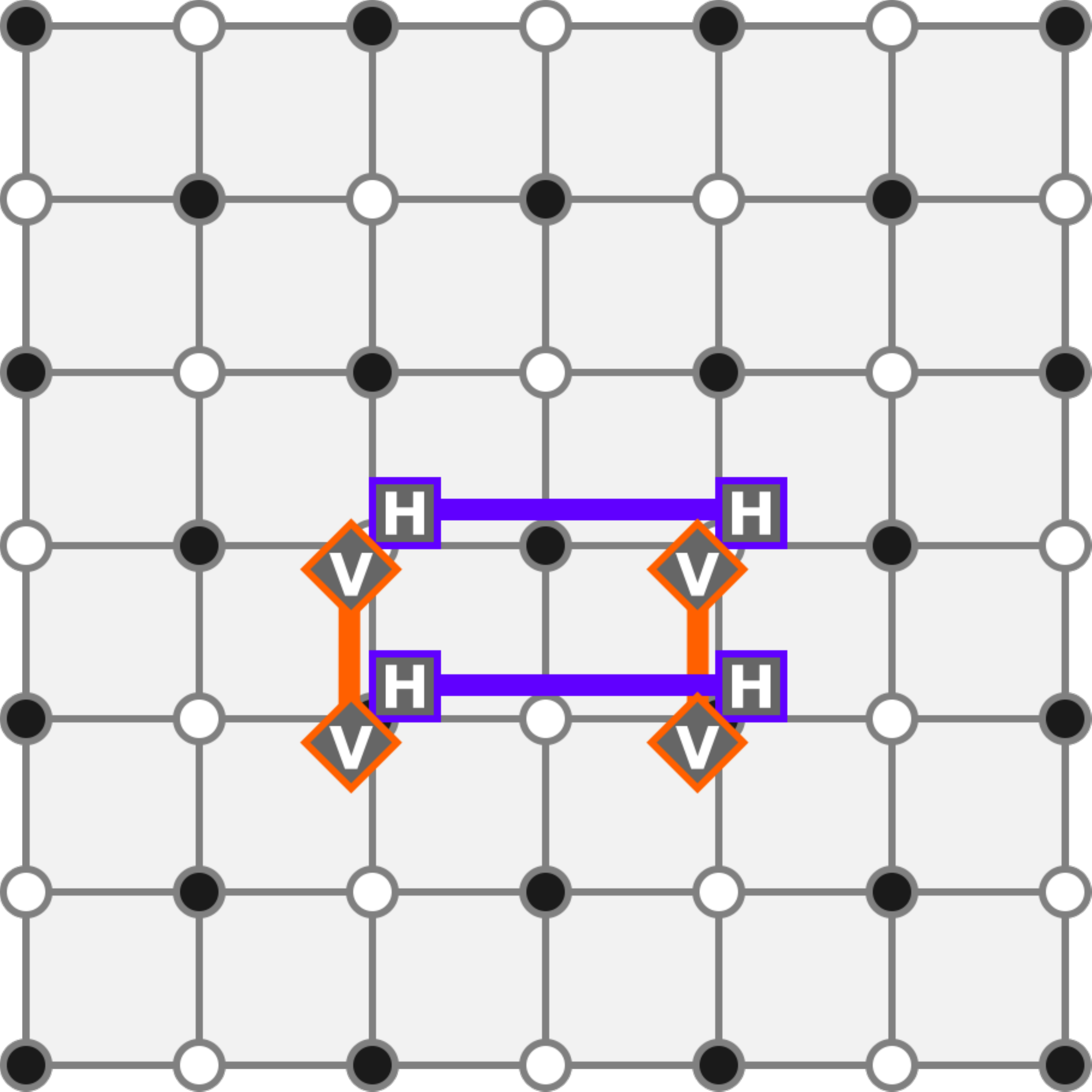}}
  \hfill
  \subfloat[]{\includegraphics[width=0.19\textwidth]{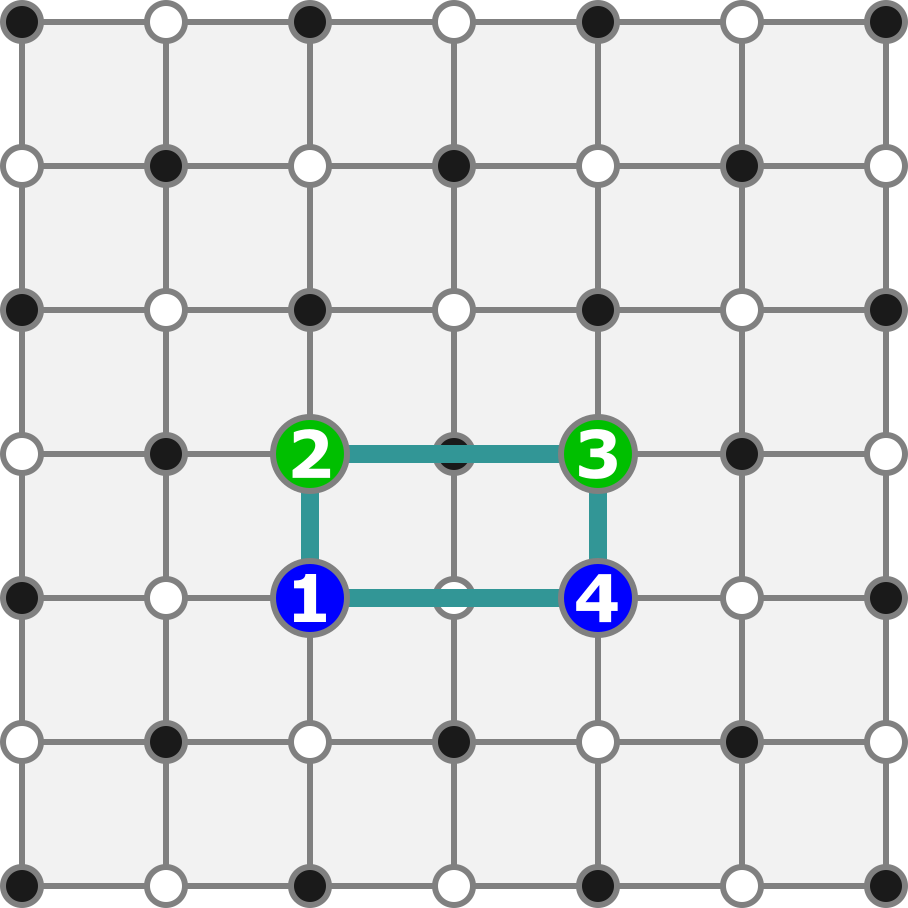}}
  \hfill
  \subfloat[]{\includegraphics[width=0.19\textwidth]{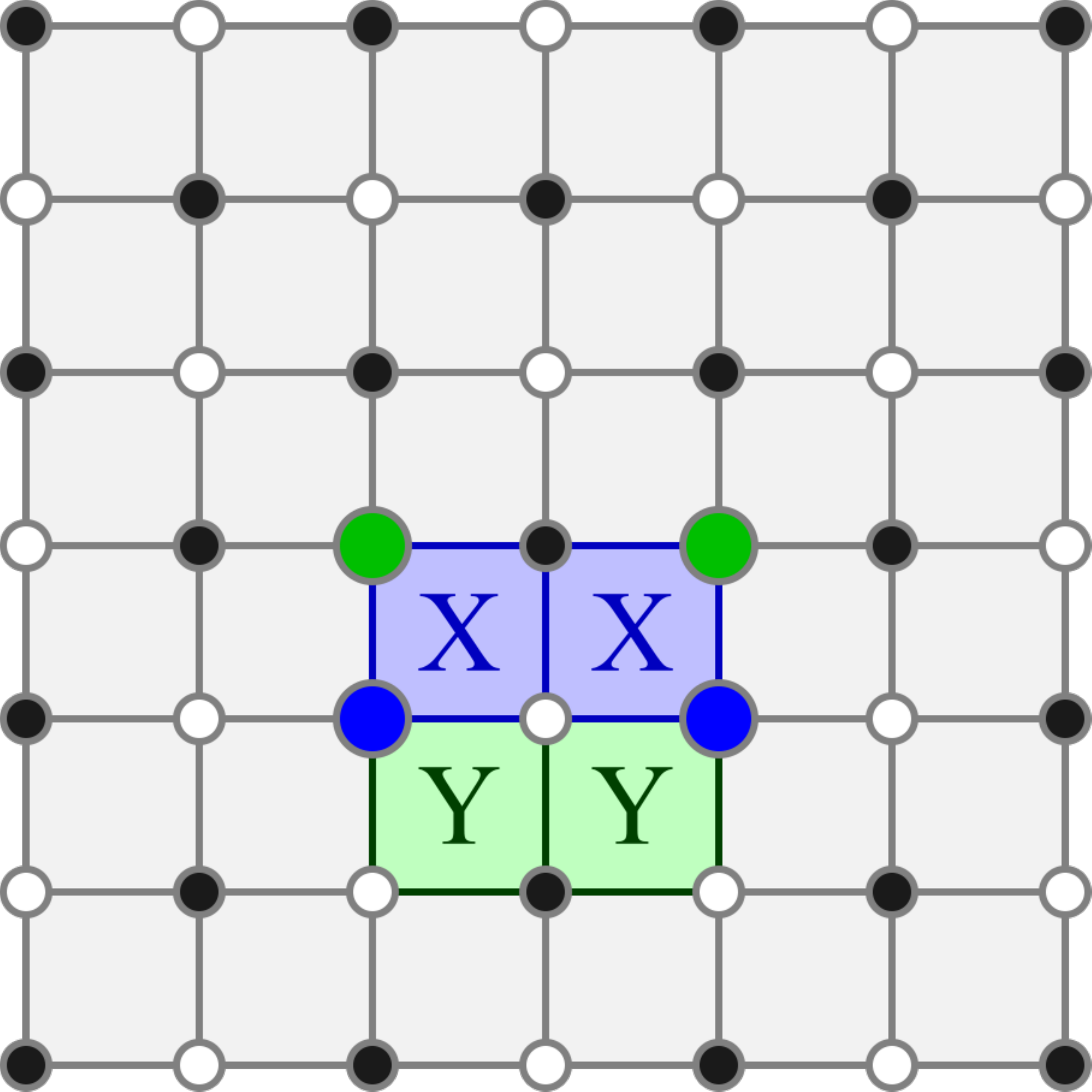}}
  \caption{Example of successful correction of an infinite bias (i.e.\ pure $Z$) error on a periodic lattice.
  (a) Error and corresponding syndrome defects.
  (b) Vertical/Horizontal graph nodes; both V and H nodes are added for each syndrome defect.
  (c) Matching from MWPM; matching is allowed between similarly oriented nodes.
  (d) Cluster of defects from following matched pairs of nodes.
  (e) Recovery operator from fusing defects around cluster; the recovery is equivalent to the original error, up to a stabilizer, and so successfully corrects the error.
  }
  \label{f-fig:6x6-simple-error}
\end{figure*}

\begin{figure*}[tb]
  \centering
  \subfloat[]{\includegraphics[width=0.194\textwidth]{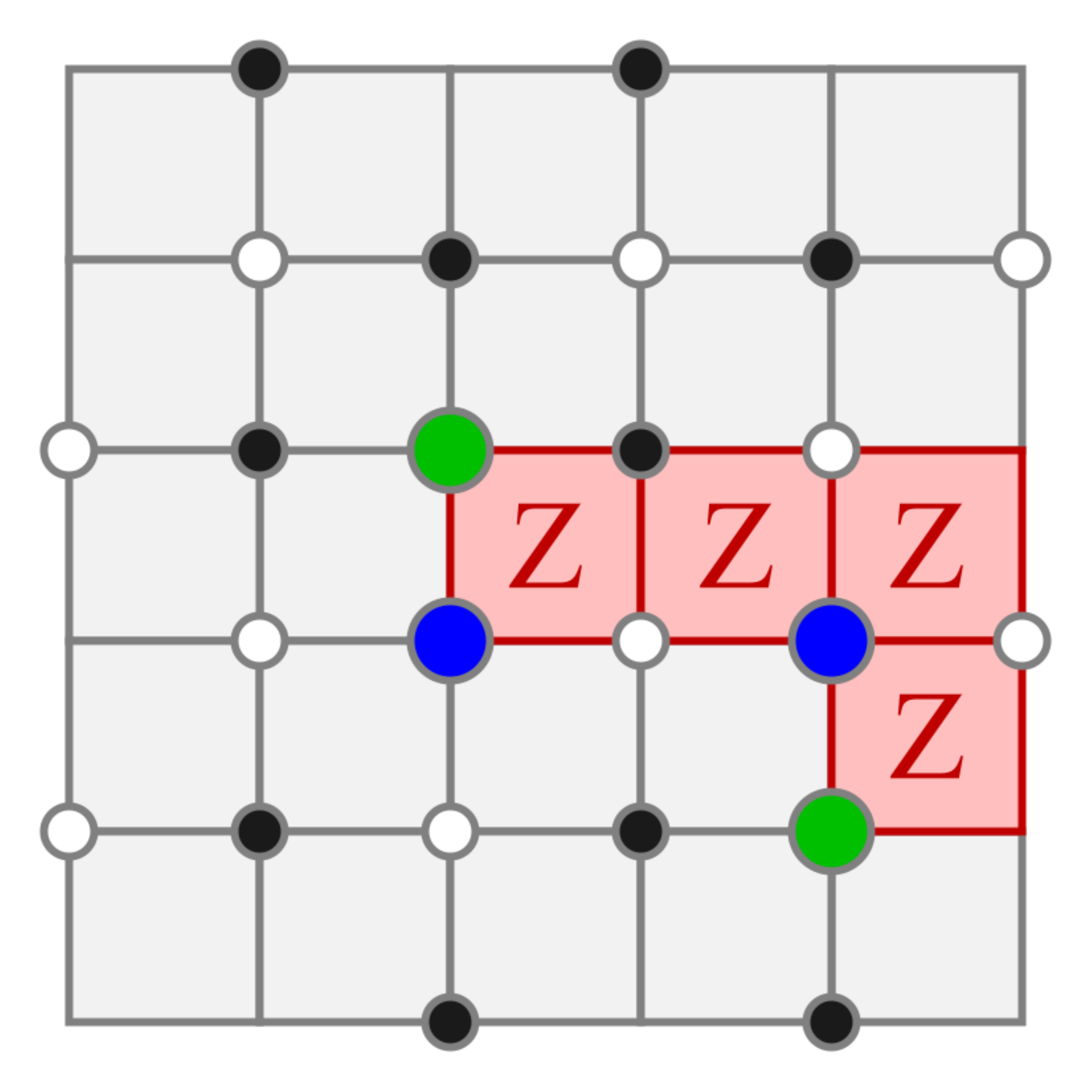}}
  \hfill
  \subfloat[]{\includegraphics[width=0.194\textwidth]{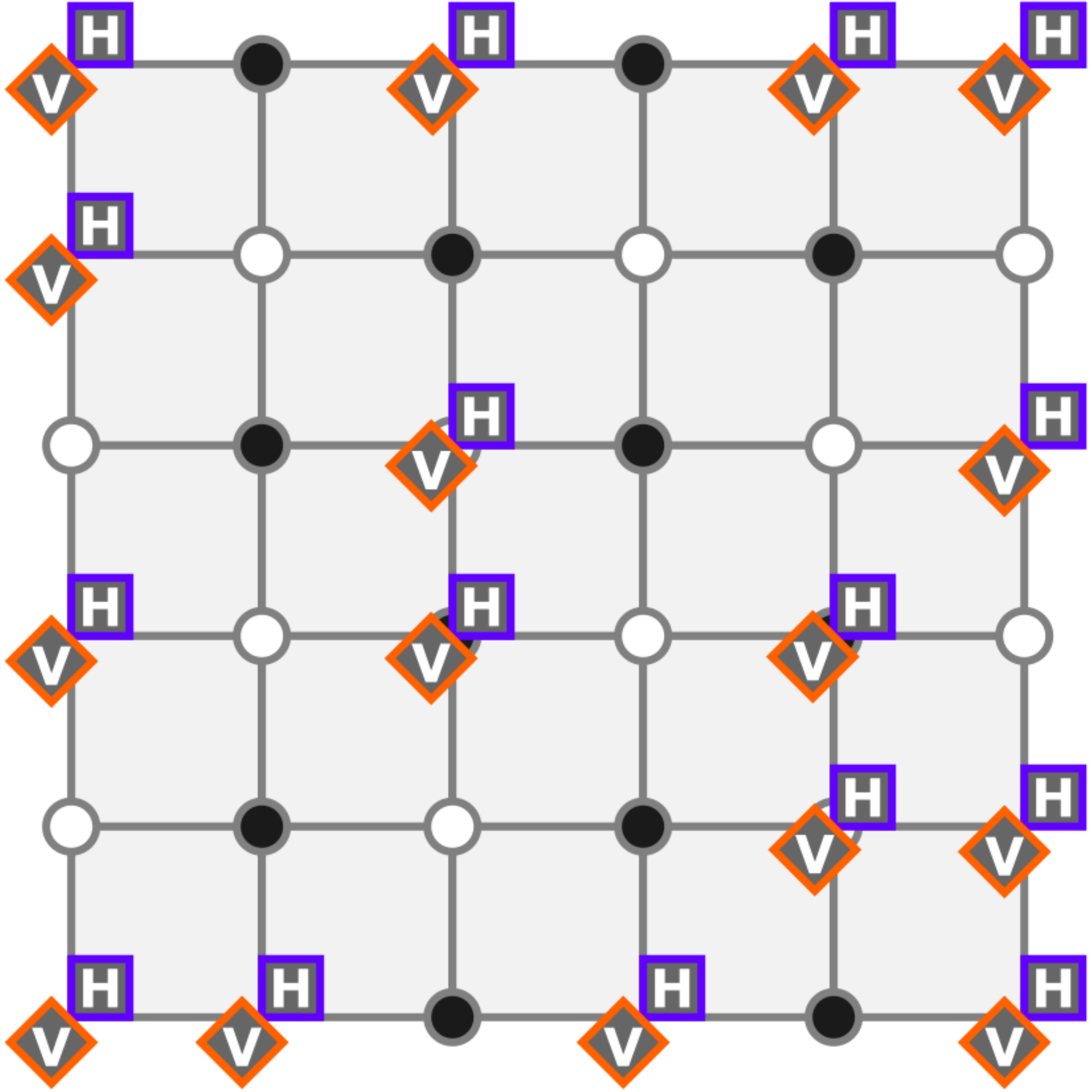}}
  \hfill
  \subfloat[]{\includegraphics[width=0.194\textwidth]{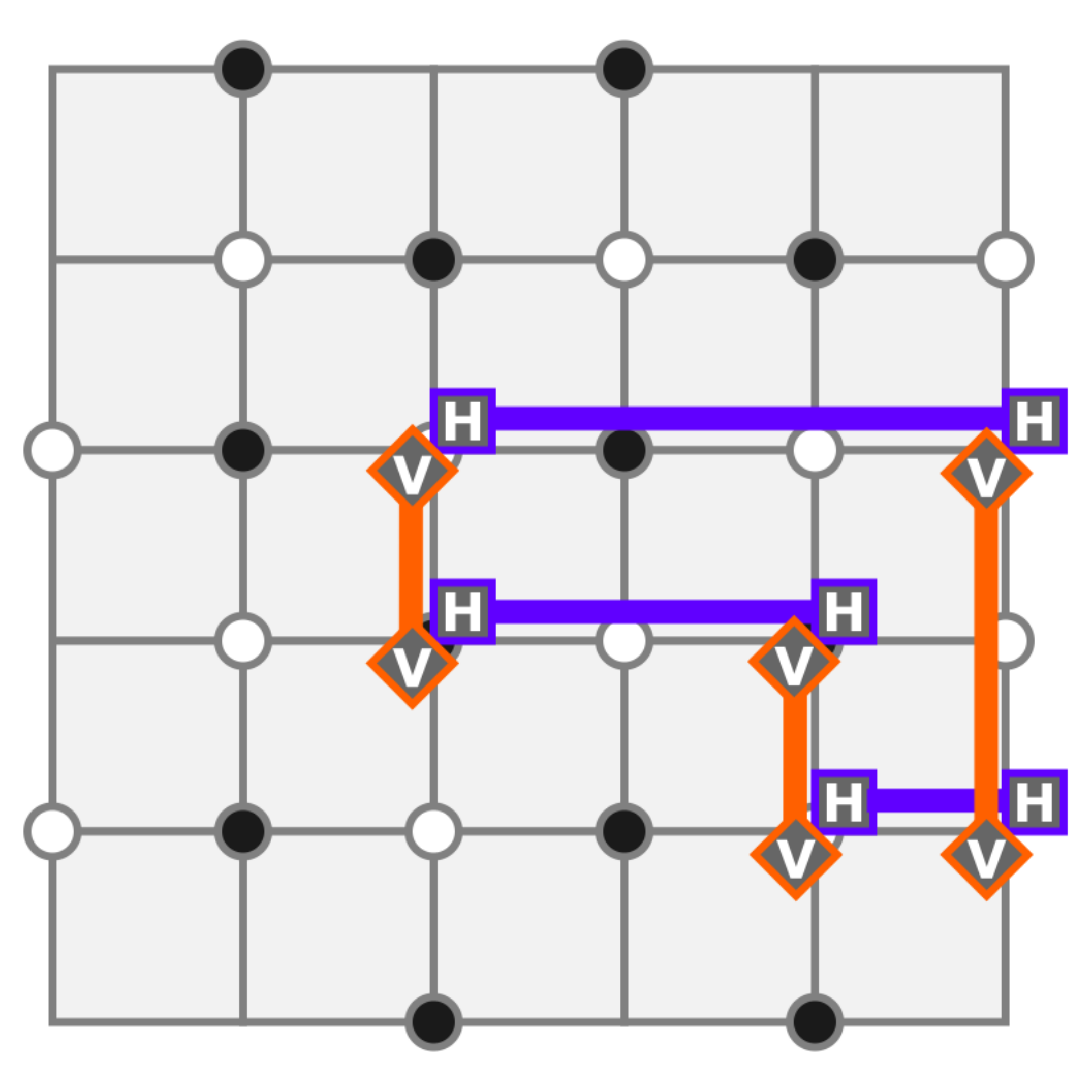}}
  \hfill
  \subfloat[]{\includegraphics[width=0.194\textwidth]{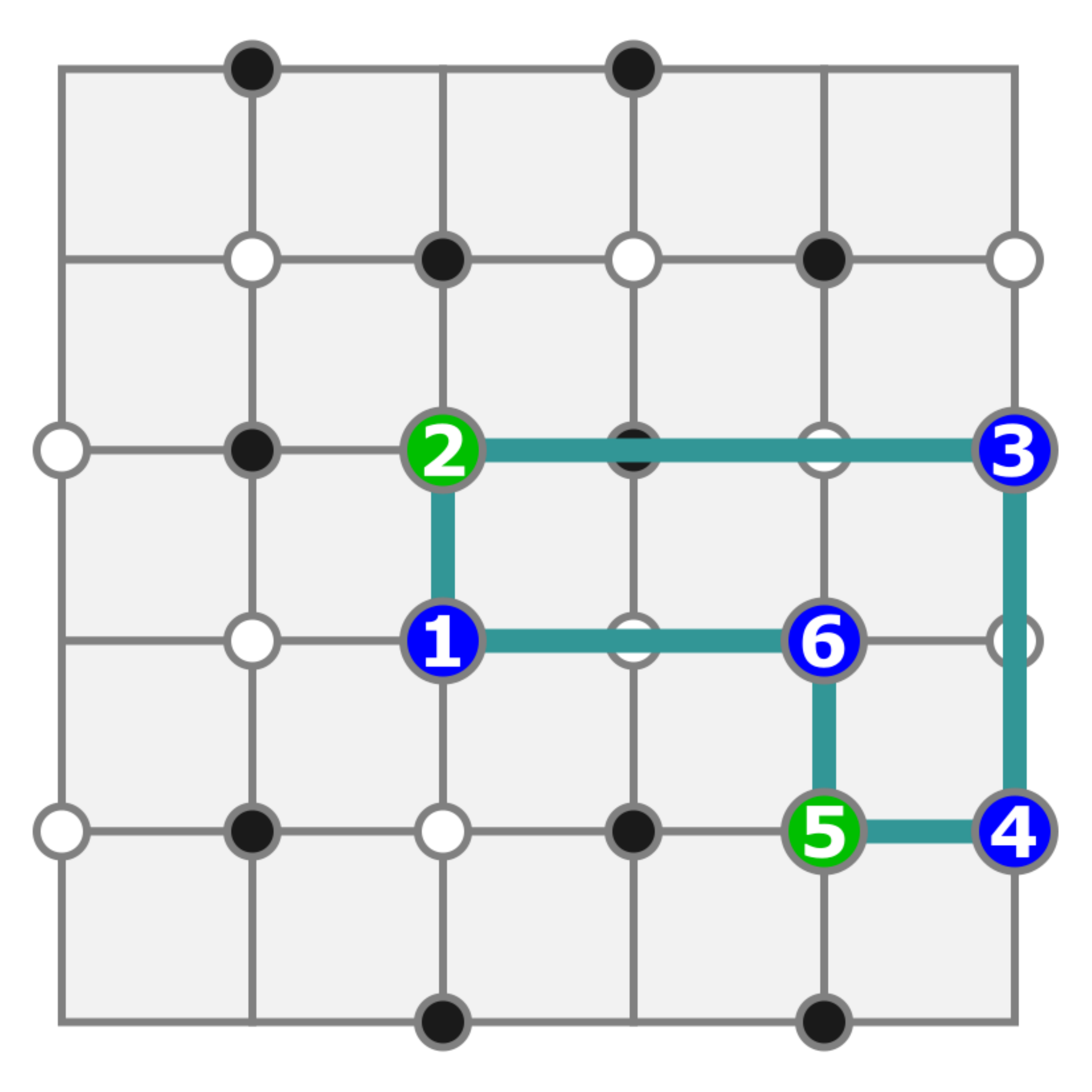}}
  \hfill
  \subfloat[]{\includegraphics[width=0.194\textwidth]{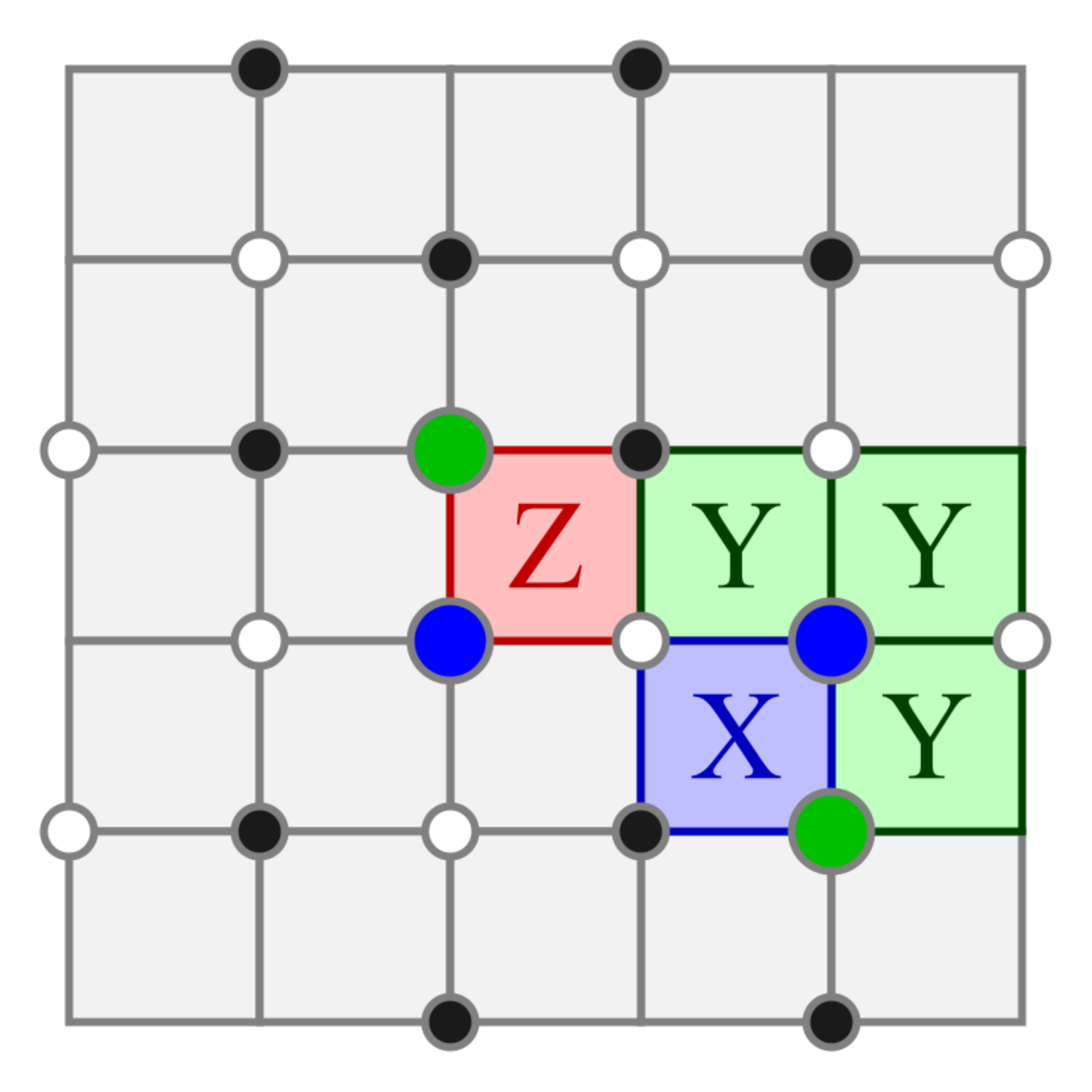}}
  \caption{Example of successful correction of an infinite bias (i.e.\ pure $Z$) error on a lattice with boundaries.
  (a) Error and corresponding syndrome defects.
  (b) Vertical/Horizontal graph defects; both V and H nodes are added for each syndrome defect, as well as virtual V and H nodes for each boundary vertex where a stabilizer is not applied.
  (c) Matching from MWPM; matching is allowed between similarly oriented nodes, and between virtual nodes at the same location (not shown here since such matchings are not used in the construction of clusters).
  (d) Cluster of defects from following matched pairs of nodes.
  (e) Recovery operator from fusing defects around cluster; the recovery is equivalent to the original error, up to a stabilizer, and so successfully corrects the error.
  }
  \label{f-fig:5x5-boundary-error}
\end{figure*}

\begin{figure*}[tb]
  \centering
  \subfloat[]{\includegraphics[width=0.19\textwidth]{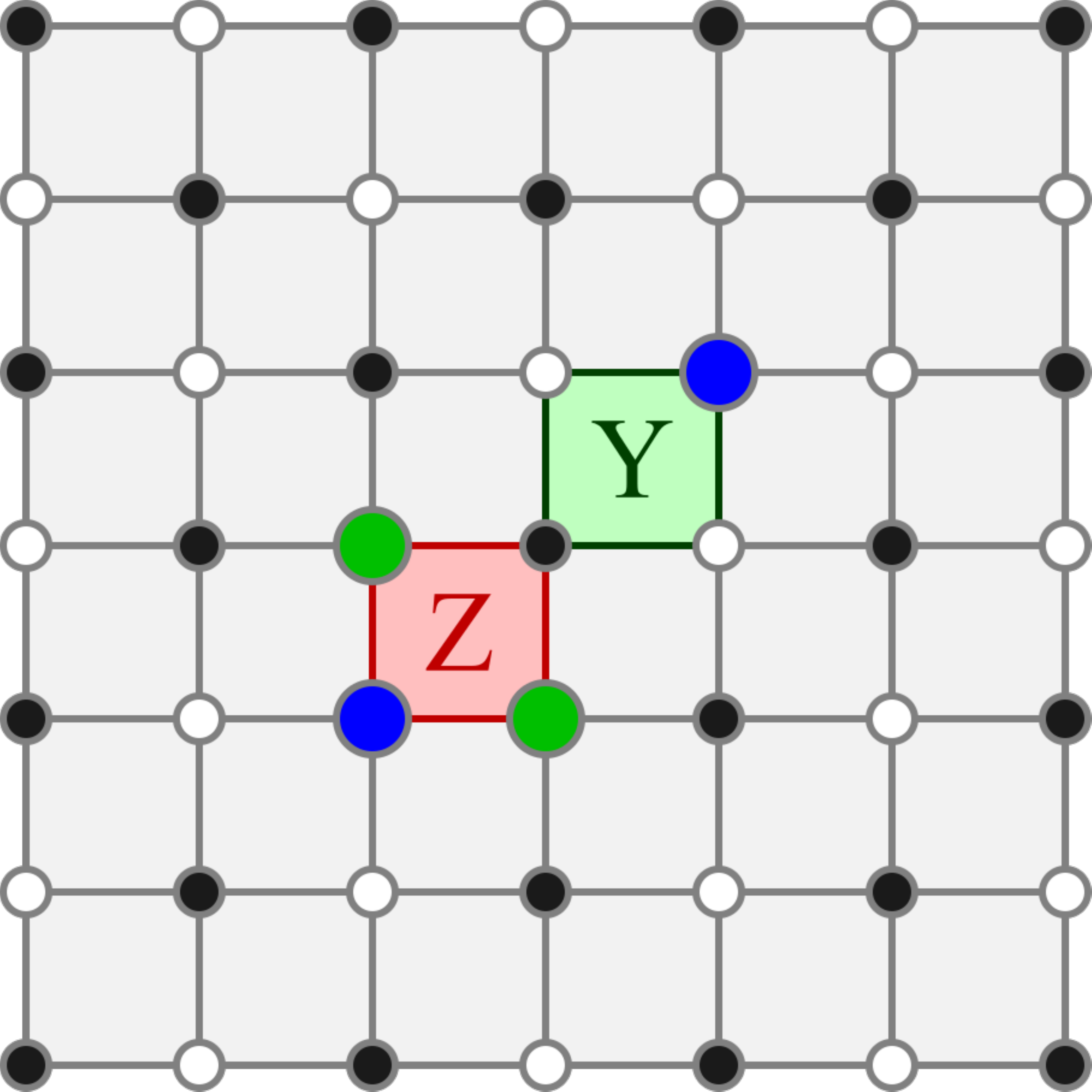}}
  \hfill
  \subfloat[]{\includegraphics[width=0.19\textwidth]{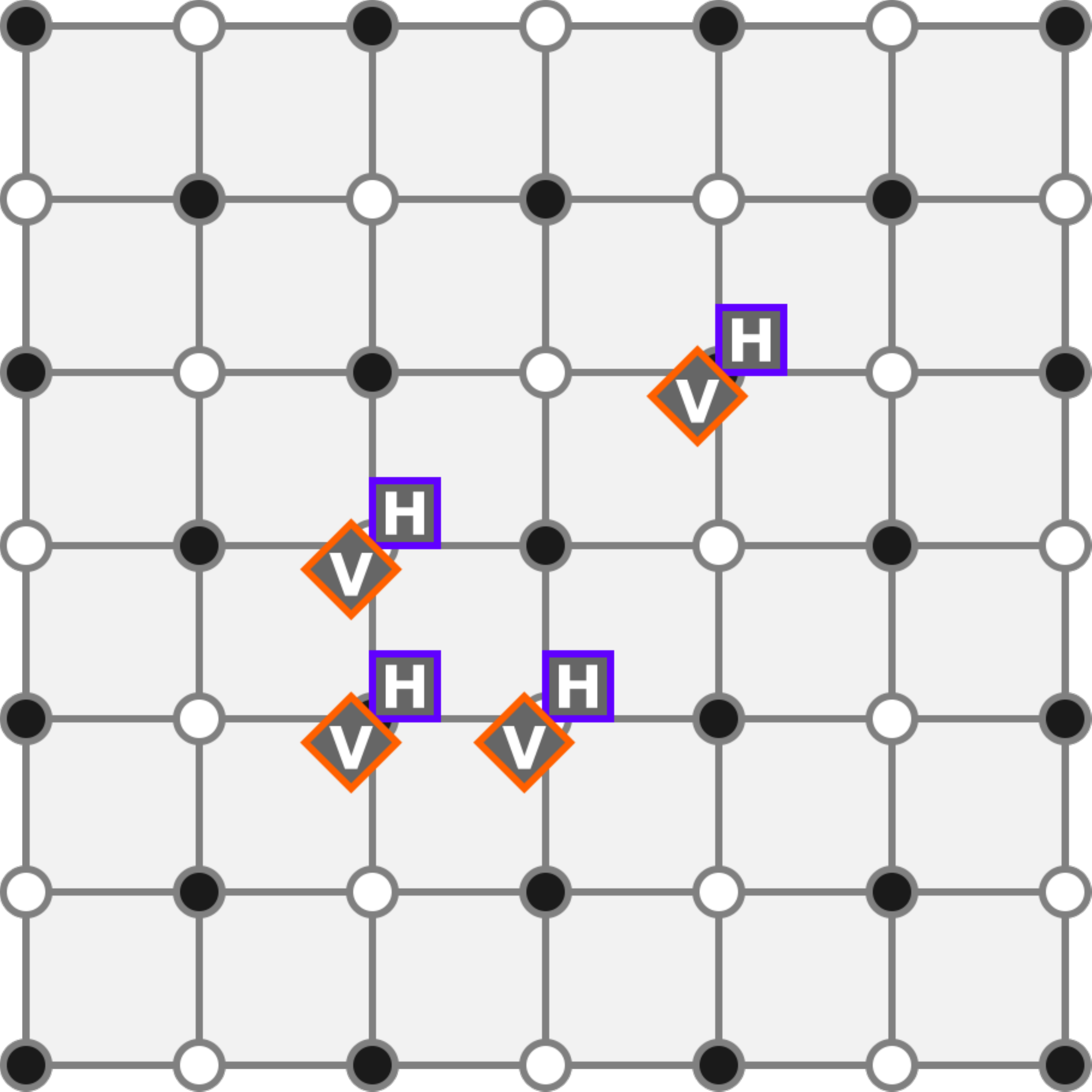}}
  \hfill
  \subfloat[]{\includegraphics[width=0.19\textwidth]{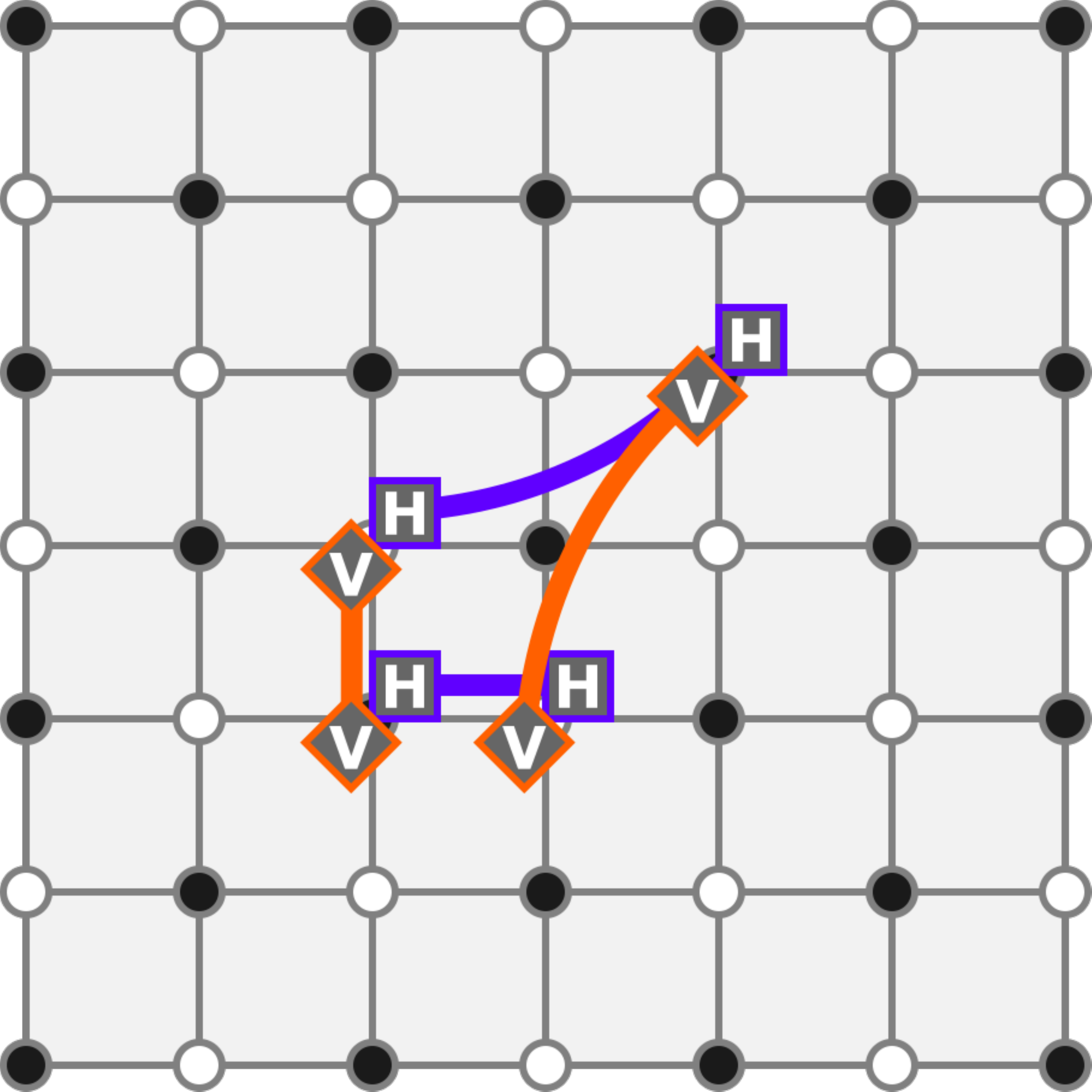}}
  \hfill
  \subfloat[]{\includegraphics[width=0.19\textwidth]{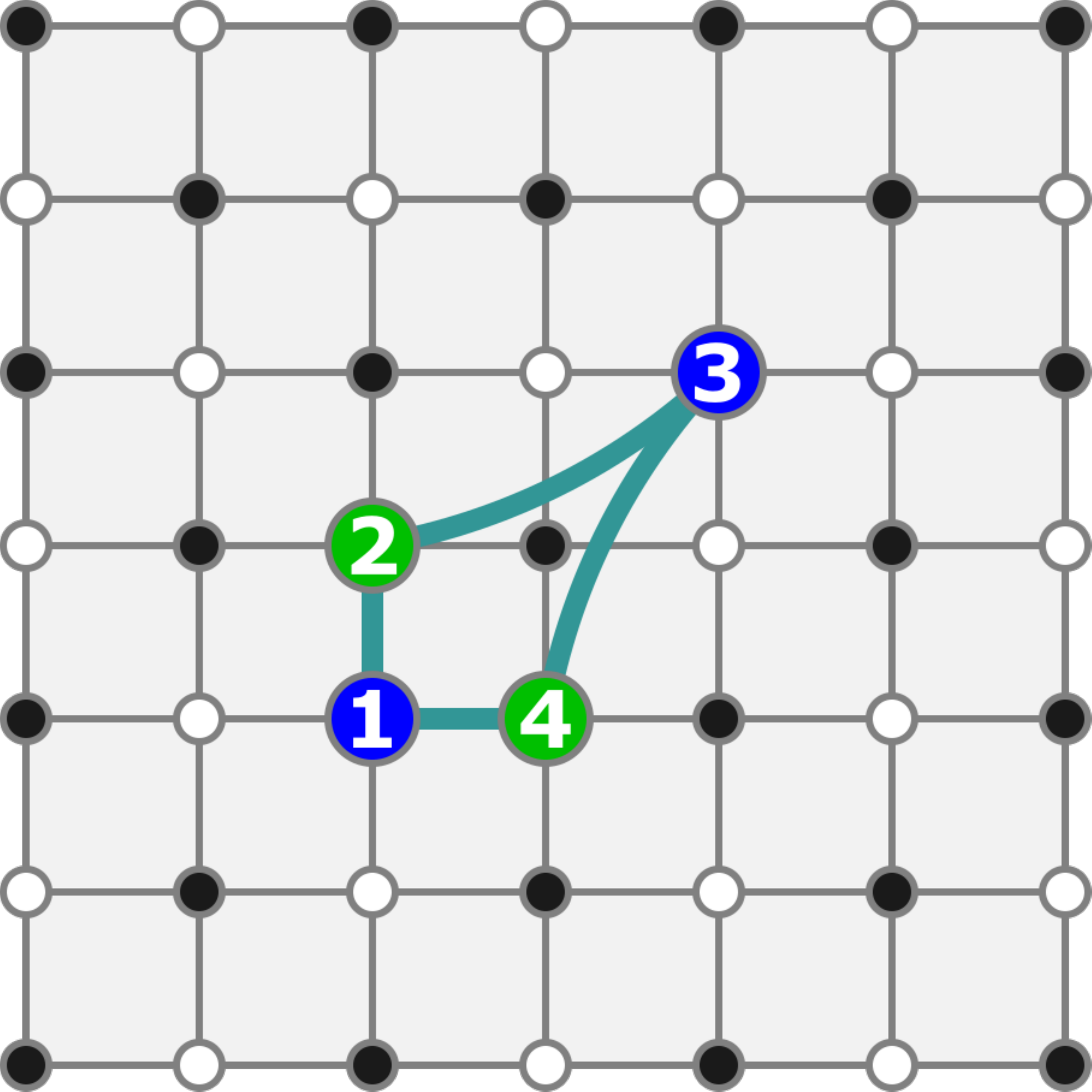}}
  \hfill
  \subfloat[]{\includegraphics[width=0.19\textwidth]{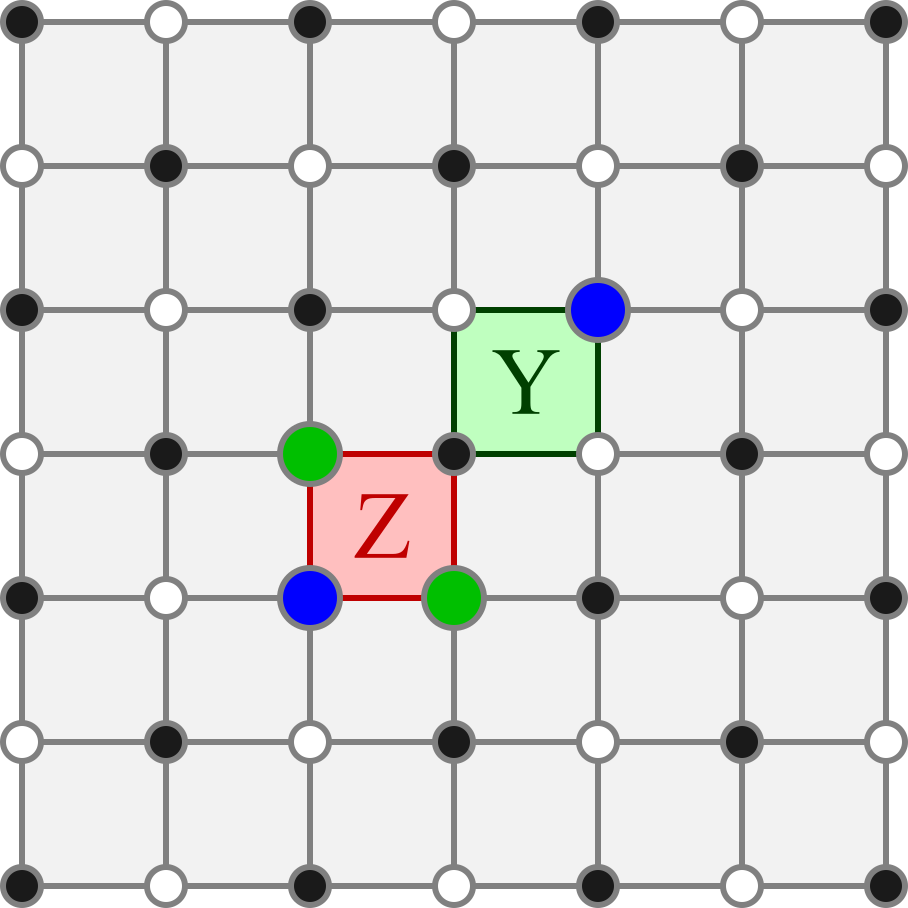}}
  \caption{Example of successful correction of a finite bias (i.e.\ high-rate $Z$ and low-rate $X$ or $Y$) error on a periodic lattice.
  (a) Error and corresponding syndrome defects.
  (b) Vertical/Horizontal graph defects; both V and H nodes are added for each syndrome defect, as well as virtual V and H nodes for each vertex where a stabilizer is not applied.
  (c) Matching from MWPM; matching is allowed between similarly oriented nodes but penalized between distinct rows or columns as a function of the noise parameters.
  (d) Cluster of defects from following matched pairs of nodes.
  (e) Recovery operator from fusing defects around cluster; the recovery successfully corrects the original error.
  }
  \label{f-fig:6x6-biased-error}
\end{figure*}

Examples showing the operation of the main decoding algorithm, assuming reliable measurements, can be seen in Figures~\ref{f-fig:6x6-simple-error}, \ref{f-fig:5x5-boundary-error} and \ref{f-fig:6x6-biased-error}.
Each figure shows: (a) an error and corresponding syndrome defects; (b) vertical/horizontal nodes added to the graph for the given syndrome defects; (c) matching of nodes resulting from MWPM over the graph; (d) cluster of defects corresponding to the matching; and (e) recovery operator resulting from neutralizing the defects in the cluster.
Figure~\ref{f-fig:6x6-simple-error} shows the successful correction of an infinite bias error on a periodic lattice.
Figure~\ref{f-fig:5x5-boundary-error} shows the successful correction of an infinite bias error on a lattice with boundaries.
Figure~\ref{f-fig:6x6-biased-error} shows the successful correction of an error due to finite bias noise on a periodic lattice.

\subsection{Decoder runtime}

The runtime of the decoder is determined by the runtime of the pairing subroutine.
With our choice of minimum-weight perfect matching, a single subroutine has runtime $O(v^3)$, where $v$ is the number of vertices of the input graph.
On average, in the main decoding step, we have $v \sim 8 pd^2$ for the ideal-measurement case and $v \sim 8 pd^3$ for the fault-tolerant case, where $p$ is the error rate and $d$ is the code distance.
We include a constant prefactor $8 = 2 \times 4$; the factor $4$ is included because each error introduces at most four defects, and the factor $2$ is included because for each defect we introduce two vertices to the graph.
We neglect the vertices that allow us to pair defects to the boundary; for large system sizes, this contribution is negligibly small compared to the number of vertices that appear in the bulk. 

In the finite-bias case, we use minimum-weight perfect matching again to pair charged clusters in a residual decoding step.
This also scales like $O(d^2)$ and $O(d^3)$ for the ideal-measurement and fault-tolerant cases, respectively.
However, in practice, this residual step is very fast compared to the main decoding step.

We note that we can choose any pairing subroutine to combine defects.
For instance, we could use the union-find decoder due to Delfosse and Nickerson~\cite{Delfosse17} to reduce the runtime from $O(v^3)$ to almost $v$.

With periodic boundary conditions and at infinite noise bias, decoding is readily parallelized into several two-dimensional pairing subroutines~\cite{Brown19}.
Moreover, in the infinite bias case, the residual decoding step is not required.
We can therefore reduce the single pairing problem of $v \sim 8 pd^3$ vertices in the fault-tolerant case into $2d$ subroutines of $v \sim 2 p d^2$ vertices.
One may consider ways to parallelize the decoder in settings where we include boundaries or when the noise bias is finite.
We leave this to future work.

\begin{figure*}[tb]
\begin{algorithm}[H]
\SetAlgoVlined
\SetCommentSty{emph}
\SetKwData{Syndrome}{syndrome}
\SetKwData{Graph}{graph}
\SetKwData{Matching}{matching}
\SetKwData{Recovery}{recovery}
\SetKwData{Clusters}{clusters}
\SetKwData{Cluster}{cluster}
\SetKwData{DefectA}{$\alpha$}
\SetKwData{DefectB}{$\beta$}
\SetKwData{DefectC}{$\gamma$}
\SetKwFunction{Decode}{decode}
\SetKwFunction{Distance}{distance}
\SetKwFunction{MWPM}{MWPM}
\SetKwFunction{ResidualDecode}{residual-decode}
\SetKwProg{Fn}{Function}{:}{}
\Fn{\Decode}{
  \tcp{syndrome is a set of defects where a defect is a (vertex location, type) tuple}
  \tcp{noise parameters are bias, qubit and measurement error rates, respectively}
  \KwIn{\Syndrome, noise parameters: $\eta$, $p$, $q$}
  \tcp{recovery is set of Pauli operators with qubit locations}
  \KwOut{\Recovery}
  \Begin(construct \Graph){
    \ForEach{defect in \Syndrome}{
      add vertical and horizontal nodes to \Graph\;
    }
    \ForEach{boundary vertex at each time step where no stabilizer is applied}{
      add vertical and horizontal virtual nodes to \Graph connected by a zero weight edge\;
    }
    \ForEach{vertical node pair, horizontal node pair in \Graph}{
      add edge to \Graph weighted by \Distance{node pair, $\eta$, $p$, $q$}\;
    }
  }
  find \Matching from \Graph using \MWPM\;
  \Begin(construct \Clusters){
    \tcp{clusters is a set of clusters where a cluster is an ordered list of defects}
    remove matches between virtual nodes with same location from \Matching\;
    \tcp{all matches are now pairs of vertical or horizontal nodes}
    \While{\Matching not empty}{
      \Begin(construct \Cluster){
        let \DefectA reference any vertical node in \Matching\;
        \Repeat{select from \Matching fails}{
          select and remove vertical node pair (\DefectA, \DefectB) from \Matching\;
          add defect corresponding to \DefectA to \Cluster\;
          select and remove horizontal node pair (\DefectB, \DefectC) from \Matching\;
          add defect corresponding to \DefectB to \Cluster\;
          let \DefectA reference \DefectC\; 
        }
        label \Cluster neutral (charged) if even (odd) parity of single-type defects\;
        add \Cluster to \Clusters\;
      }
    }
  }
  \Begin(construct \Recovery){
    \ForEach{\Cluster in \Clusters}{
      add shortest path of $Y$ ($X$) operators to \Recovery between successive $X$-type ($Y$-type) defect pairs in cluster (ignoring time dimension)\; 
    }
    Update \Recovery with output of \ResidualDecode{\Clusters}\;
  }
  \KwRet \Recovery\;
}
\caption{\label{f-algo:decode}
Main decoding algorithm
}
\end{algorithm}  
\end{figure*}

\begin{figure*}[tb]
\begin{algorithm}[H]
\SetAlgoVlined
\SetCommentSty{emph}
\SetKwData{DefectA}{$\alpha$}
\SetKwData{DefectB}{$\beta$}
\SetKwFunction{Distance}{distance}
\SetKwProg{Fn}{Function}{:}{}
\Fn{\Distance}{
  \tcp{node is a (vertex location, type) tuple with orientation and virtuality labels}
  \tcp{noise parameters are bias, qubit and measurement error rates, respectively}
  \KwIn{node pair (\DefectA, \DefectB), noise parameters: $\eta$, $p$, $q$}
  \tcp{weight is a real number}
  \KwOut{weight}
  \tcp{break path between nodes into time, parallel and diagonal steps}
  \tcp{parallel refers to horizontal (vertical) steps between horizontal (vertical) nodes}
  $\delta_t \leftarrow$ minimum number of time steps between \DefectA and \DefectB\;
  $\delta_p \leftarrow$ minimum number of parallel steps between \DefectA and \DefectB\;
  $\delta_d \leftarrow$ minimum number of perpendicular steps between \DefectA and \DefectB\;
  \eIf{$\delta_p \geq \delta_d$}{
    $\delta_p \leftarrow \delta_p - \delta_d$\;}{
    $\delta_p \leftarrow (\delta_d - \delta_p) \bmod 2$\;}
  \tcp{define time, parallel and diagonal step weights}
  $\mu_t = -\log[q/(1-q)]$\;
  $\mu_p = -\log[\eta/(\eta+1)]-\log[p/(1-p)]$\;
  $\mu_d = -\log[1/(2(\eta+1))]-\log[p/(1-p)]$\;
  \KwRet $\delta_t \mu_t + \delta_p \mu_p + \delta_d \mu_d$\;
}
\caption{\label{f-algo:distance}
Distance function
}
\end{algorithm}
\end{figure*}

\begin{figure*}[tb]
\begin{algorithm}[H]
\SetAlgoVlined
\SetCommentSty{emph}
\SetKwData{Graph}{graph}
\SetKwData{Matching}{matching}
\SetKwData{Recovery}{recovery}
\SetKwData{Clusters}{clusters}
\SetKwData{Cluster}{cluster}
\SetKwFunction{ResidualDecode}{residual-decode}
\SetKwProg{Fn}{Function}{:}{}
\Fn{\ResidualDecode}{
  \tcp{clusters is a set of cluster where each cluster is an ordered list of defects}
  \KwIn{\Clusters}
  \tcp{recovery is set of Pauli operators with qubit locations}
  \KwOut{\Recovery}
  \Begin(construct \Graph){
    \ForEach{\Cluster in \Clusters}{
      \If{\Cluster is charged}{
        add \Cluster to \Graph\;
      }
      \If{\Cluster is neutral and contains $X$-type and $Y$-type defects}{
        add two copies of \Cluster to \Graph connected by a zero weight edge\;
      }
    }
    \ForEach{corner vertex at each time step where no stabilizer is applied}{
      add virtual \Cluster to \Graph\;
    }
    \ForEach{\Cluster pair in \Graph}{
      add edge to \Graph weighted by minimum Manhattan distance between any pairing of defects drawn one from each \Cluster\;
    }
    \If{odd number of charged \Clusters}{
      add virtual \Cluster to \Graph\ with zero weight edge to each virtual \Cluster\;
    }
  }
  find \Matching from \Graph using \MWPM\;
  \Begin(construct \Recovery){
    remove matches where each \Cluster is virtual\;
    \ForEach{\Cluster pair in \Matching}{
      add shortest path of $Y$ ($X$) operators to \Recovery between selected $X$-type ($Y$-type) defects from each \Cluster in pair\;
    }
  }
  \KwRet \Recovery\;
}
\caption{\label{f-algo:residual-decode}
Residual decoding sub-algorithm
}
\end{algorithm}
\end{figure*}

\end{document}